\pdfoutput=1
\documentclass[journal]{IEEEtran}

\usepackage{cite}
\usepackage{amsmath,amssymb,amsfonts,amsthm}
\usepackage[T1]{fontenc}
\usepackage[utf8]{inputenc}
\usepackage{babel}
\usepackage{textcomp}
\usepackage{mathtools}
\usepackage{etoolbox}
\usepackage{xcolor}
\usepackage{microtype}
\usepackage[pdftex]{graphicx}
\usepackage{threeparttable}
\usepackage{rotating}

\usepackage{booktabs}
\usepackage{caption}
\usepackage{multirow}
\usepackage{multicol}
\usepackage{subfigure}
\usepackage{wrapfig}

\usepackage{tikz}
\usepackage{pgfplots}

\usetikzlibrary{
    arrows,
    arrows.meta,
    calc,
    fit,
    patterns,
    decorations.markings,
    decorations.pathmorphing,
    decorations.pathreplacing
}

\pgfplotsset{compat=1.18}
\usepgfplotslibrary{colormaps}
\usepgfplotslibrary{colorbrewer}

\usetikzlibrary{external}
\tikzset{external/mode=graphics if exists}

\pgfplotsset{
    every axis/.style={
        scaled ticks=false,
        grid=major,
        xlabel near ticks,
        ylabel near ticks,
        legend pos=south east,
        legend style={font=\footnotesize},
        every x tick label/.append style={font=\footnotesize},
        every y tick label/.append style={font=\footnotesize},
        cycle list/Set1-6,
        cycle multiindex* list={
            mark list\nextlist
            Set1-6\nextlist
            linestyles\nextlist
        },
        every axis plot/.append style={mark size=1pt, thick},
        enlarge y limits=0.025,
        enlarge x limits=0,
        legend cell align={left}
    },
    mesh/color input=explicit,
    mesh/rows=10,
    mesh/ordering=y varies,
    unbounded coords=jump,
    clip=true,
    max space between ticks=50pt
}

\usepackage{algorithm}
\usepackage{comment}
\usepackage{glossaries-extra}
\usepackage{siunitx}
\usepackage{algpseudocodex}
\usepackage{svg}
\usepackage{balance}
\usepackage{soul}
\usepackage{url}
\usepackage{nicefrac}
\usepackage{listings}
\usepackage[subtle]{savetrees}
\usepackage[none]{hyphenat}

\usepackage[pdfencoding=auto,psdextra]{hyperref}
\usepackage[capitalise]{cleveref}

\graphicspath{{figures/}}

\svgsetup{inkscapelatex=false}

\newtheorem{remark}{Remark}

\newtheorem{definition}{Definition}

\makeatletter
\newcommand\fs@betterruled{%
  \def\@fs@cfont{\bfseries}\let\@fs@capt\floatc@ruled
  \def\@fs@pre{\vspace*{5pt}\hrule height.8pt depth0pt \kern2pt}%
  \def\@fs@post{\kern2pt\hrule\relax}%
  \def\@fs@mid{\kern2pt\hrule\kern2pt}%
  \let\@fs@iftopcapt\iftrue}
\floatstyle{betterruled}
\restylefloat{algorithm}
\makeatother

\makeatletter
\renewcommand{\@thesubfigure}{\hskip\subfiglabelskip}
\makeatother

\setabbreviationstyle[acronym]{long-short}

\renewcommand{\vec}[1]{\mathbf{#1}}
\newcommand{\vecs}[1]{\boldsymbol{#1}}

\newcommand{\dv}{\vec{d}}

\newcommand{\hv}{\vec{h}}

\newcommand{\nv}{\vec{n}}

\newcommand{\pv}{\vec{p}}

\newcommand{\tv}{\vec{t}}

\newcommand{\xv}{\vec{x}}
\newcommand{\yv}{\vec{y}}
\newcommand{\zv}{\vec{z}}

\newcommand{\Id}{\vec{I}}

\newcommand{\Km}{\vec{K}}

\newcommand{\Mm}{\vec{M}}

\newcommand{\Pm}{\vec{P}}
\newcommand{\Qm}{\vec{Q}}
\newcommand{\Rm}{\vec{R}}

\newcommand{\Tm}{\vec{T}}

\newcommand{\Thetav}{\vecs{\Theta}}
\newcommand{\Phiv}{\vecs{\Phi}}

\newcommand{\Dc}{{\cal D}}

\newcommand{\Lc}{{\cal L}}

\newcommand{\Nc}{{\cal N}}
\newcommand{\Oc}{{\cal O}}
\newcommand{\Pc}{{\cal P}}

\newcommand{\CC}{\mathbb{C}}

\newcommand{\NN}{\mathbb{N}}
\newcommand{\RR}{\mathbb{R}}
\newcommand{\II}{\mathbb{I}}
\newcommand{\ZZ}{\mathbb{Z}}

\newcommand{\LB}{\left(}
\newcommand{\RB}{\right)}
\newcommand{\LP}{\left\{}
\newcommand{\RP}{\right\}}
\newcommand{\LSB}{\left[}
\newcommand{\RSB}{\right]}

\newcommand\norm[1]{\left\lVert#1\right\rVert}

\newcommand{\card}[1]{\vert{#1}\vert}
\newcommand{\logn}[2]{\mathop{\mathrm{log}_{#1} \LB #2\RB}}

\newcommand{\Pavg}{P_\mathrm{avg}}

\newcommand{\Dctrain}{\Dc_\mathrm{train}}
\newcommand{\Dcval}{\Dc_\mathrm{val}}
\newcommand{\Dctest}{\Dc_\mathrm{test}}

\newcommand{\removed}[1]{}%

\newcommand{\modified}[1]{#1}

\newcommand{\changed}[1]{#1}

\newacronym{ACM}{ACM}{adaptive coding and modulation}
\newacronym{ADC}{ADC}{analog-to-digital conversion}
\newacronym{AGC}{AGC}{automatic gain control}
\newacronym{AWGN}{AWGN}{additive white Gaussian noise}
\newacronym{BER}{BER}{bit error rate}
\newacronym{BLER}{BLER}{block error rate}
\newacronym{BP}{BP}{backpropagation}
\newacronym{BPTT}{BPTT}{backpropagation through time}
\newacronym{CE}{CE}{cross-entropy}
\newacronym{CFO}{CFO}{carrier frequency offset}
\newacronym{CSI}{CSI}{channel state information}
\newacronym{DAC}{DAC}{digital-to-analog conversion}
\newacronym{DL}{DL}{deep learning}
\newacronym{DFT}{DFT}{discrete Fourier transform}
\newacronym{FFT}{FFT}{fast Fourier transform}
\newacronym{GAN}{GAN}{generative adversarial network}
\newacronym{GRU}{GRU}{gated recurrent unit}
\newacronym{iid}{i.i.d.\@}{independent and identically distributed}
\newacronym{IFFT}{IFFT}{inverse fast Fourier transform}
\newacronym{KL}{KL}{Kullback-Leibler}
\newacronym{LSTM}{LSTM}{long short-term memory}
\newacronym{MDP}{MDP}{Markov decision process}
\newacronym{ML}{ML}{machine learning}
\newacronym{MLP}{MLP}{multilayer perceptron}
\newacronym{MIMO}{MIMO}{multiple-input multiple-output}
\newacronym{MSE}{MSE}{mean squared error}
\newacronym{NN}{NN}{neural network}
\newacronym{DNN}{DNN}{deep neural network}
\newacronym{OFDM}{OFDM}{orthogonal frequency-division multiplexing}
\newacronym{pdf}{pdf}{probability density function}
\newacronym{pmf}{pmf}{probability mass function}
\newacronym{PSNR}{PSNR}{peak signal to noise ratio}
\newacronym{RBF}{RBF}{Rayleigh block-fading}
\newacronym{ReLU}{ReLU}{rectified linear unit}
\newacronym{RL}{RL}{reinforcement learning}
\newacronym{RNN}{RNN}{recurrent neural network}
\newacronym{SFO}{SFO}{sampling frequency offset}
\newacronym{SNR}{SNR}{signal-to-noise ratio}
\newacronym{SINR}{SINR}{signal-to-interference-plus-noise ratio}
\newacronym{SGD}{SGD}{stochastic gradient descent}
\newacronym{wrt}{w.r.t.\@}{with respect to}

\newacronym{OAC}{OAC}{over-the-air computation}
\newacronym{MAC}{MAC}{multiple access channel}
\newacronym{SIC}{SIC}{successive interference cancellation}
\newacronym{TDMA}{TDMA}{time division multiple access}
\newacronym{NOMA}{NOMA}{non-orthogonal multiple access}
\newacronym{CL}{CL}{curriculum learning}
\newacronym{JSCC}{JSCC}{joint source-channel coding}
\newacronym{DeepJSCC}{DeepJSCC}{DeepJSCC}
\newacronym{MTL}{MTL}{multi-task learning}
\newacronym{MIL}{MIL}{multi-instance learning}
\newacronym{DML}{DML}{deep metric learning}
\newacronym{IoT}{IoT}{Internet of Things}
\newacronym{SSIM}{SSIM}{structural similarity index measure}
\newacronym{MS-SSIM}{MS-SSIM}{multi-scale \gls{SSIM}}

\newacronym{DDPM}{DDPM}{denoising diffusion probabilistic models}
\newacronym{MVL}{MVL}{multi-view learning}
\newacronym{CNN}{CNN}{convolutional neural network}
\newacronym{LPIPS}{LPIPS}{learned perceptual image patch similarity}
\newacronym{BPG}{BPG}{Better Portable Graphics}
\newacronym{LDPC}{LDPC}{Low-Density Parity-Check}
\newacronym{AF}{AF}{attention feature}
\newacronym{BC}{BC}{broadcast channel}

\newcommand{\appendixConstruction}{Appendix A1}

\newcommand{\appendixDatasets}{Appendix B1}

\newcommand{\appendixFairness}{Appendix C1}
\newcommand{\appendixOtherMetrics}{Appendix C3}
\newcommand{\appendixCorrelated}{Appendix C4}

\title{Learning to Interfere in Non-Orthogonal Multiple-Access Joint Source-Channel Coding}

\author{Selim F.\ Yilmaz, Can Karamanlı, Deniz Gündüz,~\IEEEmembership{Fellow,~IEEE}
\thanks{S.\ F.\ Yilmaz and C.\ Karamanli were with Department of Electrical and Electronic Engineering, Imperial College London, United Kingdom, when the work was done. S.\ F.\ Yilmaz is now with American Express Services Europe Limited, London, United Kingdom. C.\ Karamanli is now with School of Biomedical Engineering \& Imaging Sciences, King’s College London, United Kingdom. Emails: yilmazselimfirat@gmail.com, can.karamanli@kcl.ac.uk.%

D.\ Gündüz is with Department of Electrical and Electronic Engineering, Imperial College London, United Kingdom. Email: d.gunduz@imperial.ac.uk.

The present work has received funding from the European Union’s Horizon 2020 Marie Skłodowska Curie Innovative Training Network Greenedge (GA. No. 953775). We also acknowledge funding from the UKRI for the projects AI-R (ERC Consolidator Grant, EP/X030806/1) and SNS JU project 6G-GOALS under the EU’s Horizon program (Grant Agreement No. 101139232). For the purpose of open access, the authors have applied a Creative Commons Attribution (CC BY) license to any Author Accepted Manuscript version arising from this submission.}
}

\begin{document}

\maketitle

\begin{abstract}
We consider multiple transmitters aiming to communicate their source signals (e.g., images) over a multiple access channel (MAC). Conventional communication systems minimize interference by orthogonally allocating resources (time and/or bandwidth) among users, which limits their capacity. We introduce a machine learning (ML)-aided wireless image transmission method that merges compression and channel coding using a multi-view autoencoder, which allows the transmitters to use all the available channel resources simultaneously, resulting in a non-orthogonal multiple access (NOMA) scheme. The receiver must recover all the images from the received superposed signal, while also associating each image with its transmitter. Traditional ML models deal with individual samples, whereas our model allows signals from different users to interfere in order to leverage gains from NOMA under limited bandwidth and power constraints. We introduce a progressive fine-tuning algorithm that doubles the number of users at each iteration, maintaining initial performance with orthogonalized user-specific projections, which is then improved through fine-tuning steps. Remarkably, our method \modified{supports up to 64 users with potential for further scaling}, with only a 2.4\% increase in the number of trainable parameters compared to a single-user model, significantly enhancing recovered image quality and outperforming existing NOMA-based methods over a wide range of datasets, metrics, and channel conditions. Our approach paves the way for more efficient and robust multi-user communication systems, leveraging innovative ML components and strategies.
\end{abstract}
\glsresetall


\section{Introduction}
\label{sec:introduction}

\modified{In wireless communication systems, efficiently transmitting data from multiple users to a common receiver is a fundamental challenge that has driven the development of various multiple access techniques.} \modified{Traditional approaches rely on orthogonalization strategies such as} \gls{TDMA} \modified{and frequency-division multiple access (FDMA), where users are allocated dedicated time slots or frequency bands.} \Gls{TDMA} is a communication technique, where users take turns transmitting data within specific time slots. Although it enables the use of point-to-point schemes, \gls{TDMA} has lower capacity, requires timing synchronization and struggles with varying data rates~\cite{dai2015non}. \modified{In contrast,} \gls{NOMA} allows multiple users to share the same time and frequency resources simultaneously~\cite{dai2015non,verdu1998multiuser}. \modified{Power-domain \gls{NOMA} distinguishes users based on different transmission power levels, enabling the receiver to decode signals using successive interference cancellation (SIC) based on the power hierarchy.} \modified{However, in} \gls{NOMA}, signals coming from different users \modified{inevitably} interfere with each other, making the implementation \modified{significantly more complex} and computationally heavy.

\modified{The quest for improved spectral efficiency has led to increased interest in joint source-channel coding (JSCC), which combines source compression and channel coding into a unified framework.} Almost all \gls{NOMA} systems and the prior literature focus on the modulation and channel coding aspects, assuming that the source compression is carried out separately at a higher layer. Indeed, it is possible to prove the optimality of separation in the infinite block length regime when the sources at the transmitters are independent~\cite{gunduz2024joint}; therefore, most works on \gls{JSCC} over \glspl{MAC} have focused on the transmission of correlated sources~\cite{cover1980multiple, gunduz2009source, guler2018lossy,rezazadeh2019joint}. However, when it comes to practical finite block length regime, separation is suboptimal even in the case of independent inputs, and to the best of our knowledge, there are no practical joint coding algorithms that can surpass the performance of separation-based benchmarks with reasonable complexity for realistic source and channel distributions.

Recently, interest in \gls{JSCC} has been rekindled with the adoption of \glspl{DNN} for implementing \gls{JSCC}, called DeepJSCC~\cite{bourtsoulatze2019deep,xu2023deep}. This data-driven approach models the end-to-end communication system as an autoencoder architecture, enabling semantic communication that prioritizes the transmission of information that is most relevant for the underlying loss function dictated by the objective of the receiver. This relevant information is often defined as the `semantic' of the source signal. One key advantage of DeepJSCC is its ability to extract semantic information from data and map it directly to the channel input, without being limited to a finite constellation or a fixed codebook. Further developments of DeepJSCC include, but are not limited to, adaptations for various source signals~\cite{tung2022deepwive,han2022semanticpreserved}, inference tasks~\cite{jankowski}, and perceptual quality-focused image transmission~\cite{erdemir2022generative,yilmaz2024high}.

\modified{While DeepJSCC has shown remarkable success in single-user scenarios, several limitations emerge when considering multi-user networks.} The `analog' nature of DeepJSCC is critical in achieving robustness against channel variations. However, it can potentially become a limitation when it is considered in the context of a larger multi-user network. It can result in higher peak-to-average power ratio~\cite{shao2022semantic}, prevent the encryption of the transmitted messages~\cite{tung2023deep}, \modified{and} cause error accumulation over multi-hop networks~\cite{bian2024hybrid,bian2024process}. \modified{Most importantly for this work,} it does not allow decoding and removal of interference in the case of multiple interfering terminals communicating with DeepJSCC.

\begin{figure}[t!]
    \centering
    \includegraphics[width=0.9\columnwidth]{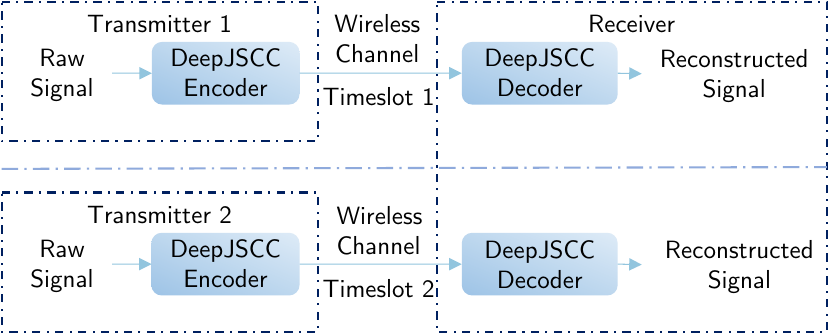}
    \caption{Extension of DeepJSCC to two users via \gls{TDMA}.}
    \label{fig:data_transmission}
\end{figure}

\modified{Extending DeepJSCC to multi-user settings presents significant challenges.} DeepJSCC can be trivially extended to multiple users via \gls{TDMA}, as shown in \cref{fig:data_transmission}. \modified{However, while} \gls{TDMA} mitigates the interference problem, it \modified{also} results in suboptimal performance compared to superposition coding that exploits \gls{NOMA}. \modified{The fundamental challenge in} extending DeepJSCC to multiple users with \gls{NOMA} \modified{stems from the fact that} different users' signals interfere with each other\modified{, creating a complex optimization problem that traditional deep learning architectures struggle to handle efficiently.} \modified{Additionally, the} inherent focus of deep learning architectures on single-sample settings with \gls{iid} data becomes problematic in multi-user scenarios, where the \gls{iid} assumption breaks down, and Shannon's separation theorem no longer holds, complicating joint encoding and decoding. \modified{This complexity has resulted in} prior work in the multi-user DeepJSCC domain \modified{being} limited to only two users \modified{due to the scalability challenges posed by} interference \modified{management and training complexity}~\cite{yilmaz2023distributed,wu2023fusion,li2023non,bo2024deep,zhang2024interference,yilmaz2024distributed}.

\modified{To address these multi-user interference challenges, we propose a novel multi-user framework that \changed{scales to many users, as demonstrated with 64 users,} by employing a multi-view autoencoder architecture with orthogonalized user-specific projections and progressive fine-tuning.} Our primary objective is to develop a JSCC scheme capable of managing multiple users, while utilizing NOMA to achieve \modified{substantial} performance improvements \modified{over traditional orthogonal approaches}. The proposed coding scheme is inspired by \modified{the} code division multiple access (CDMA) technique in digital communications, but we apply it to a continuous-amplitude modulation scheme in the context of JSCC, and learn the orthogonalization codebook employed by the users rather than using fixed chip sequences.

\subsection{Contributions}
Our main contributions are summarized as follows:
\begin{enumerate}
    \item \textbf{Novel Scalable Multi-View Autoencoder Architecture}: \modified{Unlike existing multi-user JSCC methods that use separate networks per user or fixed power allocation schemes, we} present a novel multi-view autoencoder architecture for multi-user semantic image transmission using \gls{NOMA}. We introduce \modified{learnable} user-specific projections, enabling shared encoder and decoder parameters across devices, \modified{with only 2.4\% parameter overhead for 64 users compared to existing approaches that require linear scaling,} crucial for scaling our solution to large wireless networks\footnote{We will publish the source code and model checkpoints on GitHub at \url{https://github.com/ipc-lab/deepjscc-pnoma} under the CC-BY 4.0 license.}.
    
    \item \textbf{Novel Progressive Fine-Tuning Strategy}: \modified{Addressing the training instability issues that limit existing multi-user JSCC methods to 2-3 users, we} introduce a novel progressive fine-tuning strategy that doubles the number of users at each stage, building on any point-to-point DeepJSCC scheme. It maintains performance at the start of each stage thanks to orthogonalized user-specific projections.
    
    \item \textbf{Comprehensive Performance Evaluation}: Extensive experiments show our method outperforms \gls{TDMA}-based DeepJSCC, \gls{NOMA} alternatives, and separation-based methods with BPG, neural codecs, and LDPC across all \gls{SNR} conditions for both independent and correlated source samples, ensuring fair performance among users. \modified{Our method consistently outperforms existing multi-user approaches across all SNR conditions and successfully scales to 64 users, while existing methods are limited to 2-3 users.} With 64 users, performance improves with only 2.4\% additional trainable parameters \changed{compared to a single-user architecture}.
    
    \item \textbf{Performance Analysis}: We perform comprehensive ablation studies clearly showing gains from progressive-fine-tuning, user-specific projections, and orthogonal initialization. \modified{These studies demonstrate that each component addresses specific limitations of existing multi-user JSCC approaches: progressive training solves convergence issues, learnable projections outperform fixed schemes, and orthogonal initialization ensures training stability.}
\end{enumerate}

\subsection{Organization of This Article}
The rest of the article is organized as follows. \Cref{sec:related_works} reviews related works in wireless communications, NOMA, and relevant machine learning paradigms. \Cref{sec:system_model} defines the distributed wireless image transmission problem over a MAC, outlining our objectives. \Cref{sec:methodology} introduces our methodology, including user-specific projections and a progressive fine-tuning strategy for scalable multi-user communication. \Cref{sec:numerical_results} presents numerical results, demonstrating the performance improvements of our approach compared to existing methods. \Cref{sec:conclusion} concludes the work.

\begin{table*}[t!]
\caption{Comparison of Deep Learning Based Transmission Methods for \gls{NOMA}}
\centering
\begin{tabular}{lcccccc}
\toprule
\textbf{Paper} & \begin{tabular}{@{}c@{}}\textbf{Demonstrated}\\ \textbf{Max \# Users}\end{tabular} & \textbf{Dataset(s)} & \textbf{Channel(s)} & \textbf{Coding} & \textbf{Modality} & \textbf{Uplink/Downlink} \\
\midrule
Ours & 64 & \begin{tabular}{@{}c@{}}CIFAR,TinyImagenet,\\Cityscapes,Kodak\end{tabular} & AWGN, Rayleigh & JSCC & Image & Uplink \\ \midrule
DeepJSCC-NOMA~\cite{yilmaz2023distributed} & 2 & CIFAR & AWGN & JSCC & Image & Uplink\\
DBC Aware JSCC~\cite{wu2023fusion} & 2 & CIFAR & AWGN & JSCC & Image & Downlink\\
NOMASC~\cite{li2023non} & 3 & MNIST,CIFAR,Europarl & AWGN, Rayleigh & JSCC & Image, Text & Downlink \\
DeepSCM~\cite{bo2024deep} & 2 & CIFAR & AWGN & JSCC & Image & Downlink \\
IS-SNOMA~\cite{zhang2024interference} & 2 & Cityscapes & Rician & JSCC & Image & Downlink \\
VAE-MAC~\cite{saidutta2020vae} & 2 & Gaussian & AWGN & JSCC & Bits & Downlink \\
MDC-NOMA~\cite{tang2020mdc} & 2 & Baboon,Lena & AWGN, Rayleigh & Separation-based & Image & Downlink \\
MDC-NOMA-STBC~\cite{li2022stbc} & 2 & Baboon,Lena & AWGN, Rayleigh & Separation-based & Image & Downlink \\
DeepSIC~\cite{shlezinger2020deepsic} & 2 & Gaussian & AWGN, Poisson & Separation-based & Bits & Uplink \\
CNN-SIC~\cite{sim2020deep} & 2 & Gaussian & AWGN, Rayleigh & Separation-based & Bits & Downlink \\
SICNet~\cite{van2022deep} & 2 & Gaussian & AWGN & Separation-based & Bits & Downlink \\
\bottomrule
\end{tabular}
\label{tab:comparison_noma_works}
\end{table*}

\section{Related Work}
\label{sec:related_works}

\subsection{Non-Orthogonal Multiple Access (NOMA)}

\modified{Multiple \gls{NOMA} strategies can achieve the capacity region of a \gls{MAC}, including joint decoding~\cite{cover1999elements}, \gls{SIC} combined with time-sharing (rate-splitting)~\cite{rimoldi1996rate,grant2001rate}. Our approach implements joint decoding, which can theoretically achieve the full MAC capacity region.} Recently, efforts have been made to employ \glspl{DNN} to implement \gls{SIC} for \gls{NOMA}~\cite{shlezinger2020deepsic,sim2020deep,van2022deep}. Conversely, in DeepJSCC, input signals are directly mapped to channel inputs without imposing any constellation constraints. The continuous-amplitude nature of the transmitted signals is beneficial for achieving graceful degradation with channel quality; however, it also means that the decoder functions as an estimator and will always have some noise in its reconstruction. Thus, unlike in digital communication, the decoder cannot perfectly recover the transmitted codeword, making perfect interference cancellation impossible.

\newcommand{\capacityplot}[1]{
\begin{tikzpicture}[scale=0.48]
\begin{axis}[
  xlabel = Number of users,
  ylabel = Capacity (multiple of bandwidth),
  ymin=0,
  ymax=8,
    cycle list/Set1-3,
    cycle multiindex* list={
            [3 of]mark list*\nextlist
            Set1-3\nextlist
            {solid,solid}\nextlist
    },
  mark size=3.5pt,
  title = {}]
  \addplot table[x=n,y=mac_capacity,col sep=comma] {results_other/noma_capacity.csv};
  \addplot table[x=n,y=tdma_capacity,col sep=comma] {results_other/noma_capacity.csv};
  \legend{MAC,TDMA}
\end{axis}
\end{tikzpicture}
}

\modified{Most deep learning-based NOMA and JSCC methods, such as DeepJSCC-NOMA~\cite{yilmaz2023distributed}, are limited to two-user scenarios and rely on joint decoding or deep learning-based SIC for interference management. However, perfect interference cancellation is unattainable in analog JSCC systems due to the continuous nature of the signal and inherent reconstruction noise. Our work addresses this scalability bottleneck by introducing a progressive fine-tuning algorithm and orthogonally-initialized user-specific projections, enabling efficient scaling to many users (up to 64 in our experiments) while maintaining performance.}
\Cref{tab:comparison_noma_works} presents a comparison of deep learning-based transmission methods for \gls{NOMA}. The \gls{BC} and \gls{MAC} are complementary: \Gls{BC} enables downlink transmission from one sender to multiple receivers, while \gls{MAC} supports uplink from multiple senders to one receiver. \Gls{NOMA} enhances both by allowing simultaneous transmissions, improving spectral efficiency through superposition coding in \gls{BC} and \gls{SIC} in \gls{MAC}. In our previous work~\cite{yilmaz2023distributed}, we developed a DeepJSCC scheme for distributed image transmission over a noisy \gls{MAC} using \gls{NOMA} and Siamese networks, outperforming traditional methods in low-bandwidth conditions. \modified{However, this work, like most others in the literature, considered only two users and could not be scaled to accommodate more users easily.} Our current work, however, can handle a much higher number of users with a progressive fine-tuning algorithm that effectively scales the number of users while maintaining performance through orthogonalized projections and refinement.

Tang \textit{et al.}~\cite{tang2020mdc} propose a hybrid MDC-NOMA scheme combining multiple-description coding and NOMA to enhance throughput and robustness. Li \textit{et al.}~\cite{li2022stbc} improve system reliability by applying space-time block coding (STBC) to MDC-NOMA. Cheng \textit{et al.}~\cite{cheng2024goal} introduce a goal-oriented semantic information transmission framework with message-sharing NOMA, improving efficiency by leveraging common messages among users. Zhang \textit{et al.}~\cite{zhang2024interference} present semantic difference (SeD)-aware NOMA transceivers for semantic image transmission, mitigating semantic-level interference and outperforming SeD-unaware NOMA and TDMA in both image quality and transmission efficiency. However, these studies focus exclusively on two-user scenarios, limiting their applicability to real-world communication systems accommodating more users.

Wu \textit{et al.}~\cite{wu2023fusion} propose a semantic communication system for wireless image transmission over a two-user degraded \gls{BC}. Bo \textit{et al.}~\cite{bo2024deep} present a digital semantic communication framework leveraging the hierarchical structure of semantic information for \glspl{BC} with varying channel conditions. Li \textit{et al.}~\cite{li2023non} introduce NOMASC, a NOMA-based semantic communication system for non-orthogonal transmission of text and image data, supporting only two or three users and potentially struggling with deviations from training scenarios. Saidutta \textit{et al.}~\cite{saidutta2020vae} propose a variational autoencoder achieving better reconstruction quality and robustness to channel variations. Our work surpasses similar studies by: (i) \modified{scaling up to 64 users thanks to the introduced novel techniques}, (ii) enabling efficient training via shared encoder and decoder architectures with orthogonally initialized user-specific projections, and (iii) generalizing to various channel conditions and datasets with different image sizes and domains.

\subsection{Related Deep Learning Paradigms}
\Gls{MTL} is a paradigm where multiple tasks are learned jointly to enhance the overall performance~\cite{caruana1997multitask,zhang2021survey}. In the context of \gls{MAC}, multiple signals are superimposed into a joint representation, and the decoder reconstructs multiple images, making it inherently a \gls{MTL} problem. Conventional approaches in distributed compression and JSCC~\cite{mital2022neural,yilmaz2024distributed,wu2023fusion,li2023non,bo2024deep,zhang2024interference,saidutta2020vae} typically employ distinct encoding and decoding functions for each source signal, which increases the complexity and hinders scalability. \gls{MTL} methods often modify the feature space or share parameters to improve performance across tasks\modified{~\cite{zhang2021survey}}. In the proposed scheme, inspired by these techniques, we introduce a novel multi-view autoencoder architecture for multi-user image transmission, utilizing orthogonally initialized learned projections to mitigate the adverse effects of interference during training.

Integrating multi-view information in \glspl{DNN} remains challenging. Fusion can occur at the data, feature, or output stages~\cite{mandira2019spatiotemporal,giritliouglu2021multimodal}. For example, decoder-only side information-based methods~\cite{mital2022neural,yilmaz2024distributed} merge encoded features at the decoder stage, while \gls{NOMA}-oriented approaches~\cite{yilmaz2023distributed} typically simulate \gls{MAC} by summing encoded signals at the bottleneck level. In our method, we adopt the latter strategy, combining multiple views at the bottleneck level via the channel without requiring communication between users.

Analogous to \gls{DML}, our approach maps multiple instances into a shared lower-dimensional latent space using a unified network architecture. In \gls{DML}, training typically involves a subsample of instance pairs, and the selection of these training pairs significantly influences performance~\cite{hoffer2015deep}. To address this, we have developed a systematic and efficient training methodology that circumvents the infeasibility of employing all possible training pairs.

\Gls{CL} is a progressive training paradigm where a network is initially trained on easier tasks before being adapted to the main task of interest~\cite{bengio2009curriculum}. \gls{CL} has been previously leveraged in deep learning aided channel code design in~\cite{shao2022attentioncode} by starting training at higher \glspl{SNR} and progressively adapting to lower \glspl{SNR}. In our problem, the superposition of signals leads to interference among them, adversely affecting training. To address this issue, we introduce a novel progressive fine-tuning approach that doubles the user count at each stage, building upon any point-to-point DeepJSCC system. This way, we preserve performance at the beginning of each stage through the use of orthogonal, user-specific projections.

\textbf{Notation:} Unless stated otherwise; boldface lowercase letters denote tensors (e.g., $\pv$), non-boldface letters denote scalars (e.g., $p$ or $P$), and uppercase calligraphic letters denote sets (e.g., $\Pc$). $\RR$, $\NN$, $\CC$ denote the sets of real, natural, and complex numbers, respectively. $\ZZ$ denotes the set of integers. $\card{\Pc}$ denotes the cardinality of set $\Pc$. We define $\LSB n \RSB \triangleq\{1,2,\ldots,n\}$, where $n\in\NN^+$, and $\LSB i, j \RSB \triangleq \{i, i+1,\ldots,j\}$, where $i,j \in \ZZ$ and $i < j$. We define $\II \triangleq [255]$. We define $\pv^H$ as the complex conjugate of the $u$-dimensional complex vector $\pv \in \CC^u$ for $u \in \NN^+$. $\mathbf{I}_u$ is a $u$-dimensional identity matrix for $u \in \NN^+$.

\section{Problem Definition}
\label{sec:system_model}
We consider the distributed wireless image transmission problem over a \gls{MAC} in an uplink setting with $n$ users (transmitters) and a single receiver. Let $\xv_i \in \II^{W \times H \times C_\mathrm{in}}$ denote the image of user $i$, where $W$ and $H$ denote the width and height of the image, while $C_\mathrm{in}=3$ represents the R, G and B channels for colored images. The channel is specified as $\yv = \sum_{i=1}^n h_i \zv_i + \nv$, where $\zv_i \in \CC^{k}$ is the transmitted signal vector by user i, $\nv \in \CC^{k}$ is the \gls{iid} complex Gaussian noise term with variance $\sigma^2$, i.e., $\nv \sim \CC\Nc(\vec{0}, \sigma^2 \vec{I}_{k})$, and $h_i \in \CC$ is the channel gain of the $i^\mathrm{th}$ user. We set $h_i=1, \, \forall i,$ for the \gls{AWGN} channel and $h_i \sim \CC \Nc(0, 1)$ for the Rayleigh fading channel, $\forall i \in \LSB n \RSB$. We enforce average transmission power constraints $\Pavg$ on all the users, defined by $\frac{1}{k} \norm{\zv_i}_2^2 \leq \Pavg, \, \, i \in \LSB n \RSB$. 

\modified{More explicitly, the transmit signal from user $i$ can be expressed as:
\begin{equation}
    \zv_i = E_{\Thetav_i}(\xv_i, \hv, \sigma) \in \CC^k
\end{equation}
and the received signal at the common receiver is:
\begin{equation}
    \yv = \sum_{i=1}^n h_i \zv_i + \nv = h_1\zv_1 + h_2\zv_2 + \cdots + h_n\zv_n + \nv \in \CC^k,
\end{equation}
where the superposition of all user signals creates interference that must be managed through joint decoding.}

We define the average \gls{SNR} as $
    \label{eq:snr}
    \mathrm{SNR} = 10 \logn{10}{\frac{\Pavg}{\sigma^2}} \,\, \si{\decibel}$. The {\it bandwidth ratio} $\rho$ characterizes the available channel resources per-user, and is defined as $\rho \triangleq \frac{k}{ C_\mathrm{in} W H} \  \mathrm{channel \ symbols / pixel}$. 

\modified{\textbf{Architectural Design Considerations:}
Multi-user JSCC requires balancing performance, scalability, and efficiency. Key decisions include: (1) joint vs. sequential decoding, (2) parameter sharing vs. user differentiation, (3) structured interference management, and (4) variable user count support. These motivate our multi-view autoencoder with learnable user-specific projections and progressive training.}

 Throughout this work, we assume that $\sigma$ and $\hv = \begin{bmatrix}
    h_1 & \ldots & h_n
\end{bmatrix}^T$ are known to the users and the receiver. Therefore, the $i^\mathrm{th}$ user employs a non-linear encoding function $E_{\Thetav_i}$, parameterized by $\Thetav_i$, to map its image into a complex-valued latent vector $\zv_i=E_{\Thetav_i} (\xv_i, \hv, \sigma) \in \CC^k$, where $k$ is the available channel bandwidth and $\hv$ is the vector of channel gains. A non-linear decoding function  $D_{\Phiv}$, parameterized by $\Phiv$, \modified{that implements joint decoding by processing the combined received signal from all users simultaneously to} reconstructs all the images that are aggregated in the common channel output $\yv$, to obtain $    \begin{bmatrix}
        \hat{\xv}_1 & \ldots & \hat{\xv}_n
    \end{bmatrix}^T =  D_{\Phi} (\yv,\hv,\sigma)$. \modified{This joint decoding approach, while computationally more complex than successive interference cancellation methods, enables the decoder to learn the joint distribution of all user signals and can theoretically achieve the full MAC capacity region. Our neural network implementation provides a practical approximation to the theoretically optimal joint decoding scheme, which along with joint encoding, superposition coding with SIC and time-sharing~\cite{rimoldi1996rate}, and rate-splitting approaches~\cite{rimoldi1996rate} represents one of the MAC capacity-achieving strategies.} \modified{However, joint decoding complexity scales quadratically with user count, which may become computationally prohibitive for very large numbers of users, representing a fundamental scalability limitation of this approach.} For fair comparison among methods with different number of users, we define the per-user bandwidth ratio as $\Bar{\rho} \triangleq \rho/n$ and per-user average power constraint as $\Bar{P}_\mathrm{avg} \triangleq \Pavg/n$. Throughout this work, we use a per-user average power constraint of $\Bar{P}_\mathrm{avg}=1$ for all the methods to make the TDMA-based DeepJSCC model comparable with previous works in the DeepJSCC literature~\cite{bourtsoulatze2019deep,tung2022deepjscc}.

The goal is to maximize the average \gls{PSNR}, on an unseen target dataset defined as $\mathrm{PSNR}(\xv, \hat{\xv}) = 10 \logn{10}{\frac{A^2}{
\frac{1}{C_\mathrm{in}HW}
\norm{\xv - \hat{\xv}}_2^2 }} \,\, \si{\decibel}$, where $A$ is the maximum possible input value, e.g., $A=255$ for images with 8-bit per channel. \modified{Additional evaluation metrics including SSIM, MS-SSIM, and LPIPS are defined in the appendix for comprehensive perceptual quality assessment.}

\begin{figure*}[!t]
    \centering
    \includegraphics[width=0.9\textwidth]{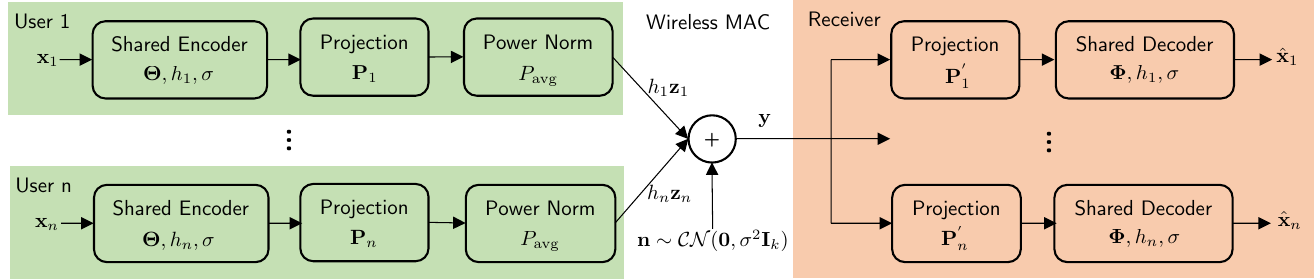}
    \caption{Overall architecture of our DeepJSCC-PNOMA method. Due to the design of the user-specific projection matrices in DeepJSCC-PNOMA—which preserve the power of each encoded signal and enforce mutual orthogonality among signals—the system with \(n\) users initially behaves as if it were time-sharing among \(\frac{n}{2}\) users under fixed per-user average power \(\Bar{P}_\mathrm{avg}\) and per-user bandwidth ratio \(\Bar{\rho}\). In this configuration, each user's signal retains its original power and remains free from interference, enabling independent decoding. Consequently, the network can begin fine-tuning without any performance degradation from inter-user interference.}
    \label{fig:architecture}
\end{figure*}

\begin{figure*}[t!]
    \centering
    \includegraphics[width=0.9\linewidth]{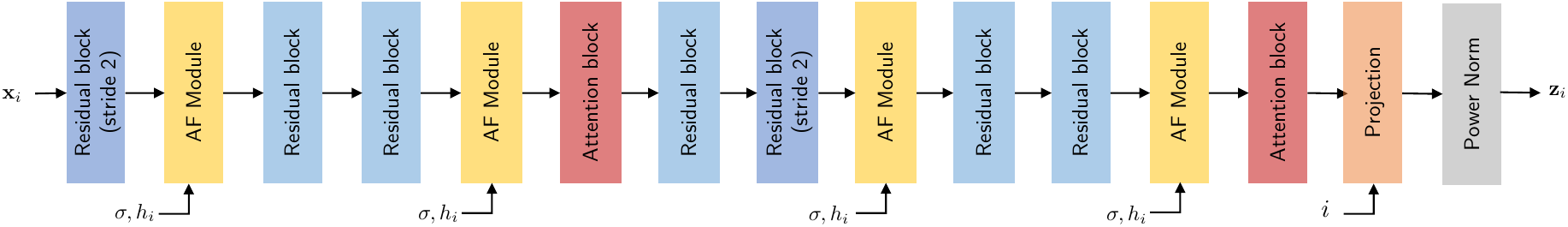}
    \caption{Encoder architecture, user-specific projection layer and power normalization layer of the introduced method.}
    \label{fig:noma_enc}
\end{figure*}

\begin{figure*}[t!]
    \centering
    \includegraphics[width=0.9\linewidth]{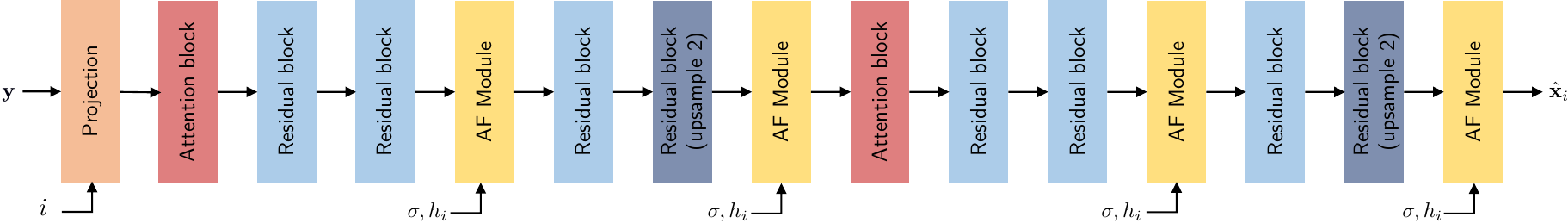}
    \caption{User-specific projection layer and decoder architecture of the introduced method.}
    \label{fig:noma_dec}
\end{figure*}

\section{Methodology}
\label{sec:methodology}

\subsection{Point-to-Point DeepJSCC Architecture}
To construct a point-to-point \gls{JSCC} system for image transmission, we utilize an autoencoder architecture based on the encoder and decoder designs from \cite{tung2022deepjscc}. This architecture features symmetric encoder and decoder networks, each with two downsampling and two upsampling layers. It incorporates residual connections and the attention mechanism introduced in \cite{cheng2020learned}, enhancing model performance. Additionally, it leverages \gls{SNR} adaptivity using the \gls{AF} module as proposed in \cite{xu2021wireless,wu2022channel}. This module allows the same model to be used during training and testing across channels with different \glspl{SNR} without significant performance degradation. By including \gls{SNR} as an input feature and training with randomly sampled \glspl{SNR}, the network learns parameters for various \gls{SNR} conditions. %

In point-to-point DeepJSCC, the transmitter encodes the input as $\Tilde{\zv}=E_{\Thetav}(\xv, h, \sigma)$, then flattens and normalizes it by dividing by $\sqrt{k \Pavg/(\zv \zv^H)}$ to obey the power constraint, where $\zv^H$ is the complex conjugate transpose of $\zv$. The receiver decodes the noisy channel output $\yv$ using $\hat{\xv}=D_{\Phiv} \LB \yv, h, \sigma \RB$.

\begin{remark}[Extensibility]
\label{remark:extensibility}
The encoder and decoder architectures are interchangeable with other autoencoders. Our components directly utilize the encoder and decoder without modifications or constraints, allowing for flexible adoption of new architectures and performance improvements.
\end{remark}

\subsection{Parameter Sharing Between Users via Novel User-Specific Projections}
\label{sec:projection}

We extend point-to-point architecture to multi-user setting over a \gls{MAC} by introducing a novel superposition coding technique that improves scalability and performance of multi-user neural networks. \Cref{fig:architecture} summarizes our method, named DeepJSCC-PNOMA. We assume that no communication occurs between users during the transmission phase. \Cref{fig:noma_enc,fig:noma_dec} illustrate the encoder and decoder architectures for our multi-user DeepJSCC, respectively. This architecture is flexible and can be substituted with any other encoder and decoder architecture, as mentioned in \cref{remark:extensibility}.

\modified{Multi-user JSCC requires managing interference while maintaining scalability. Traditional approaches use either separate networks per user (leading to 6300\% parameter overhead for 64 users) or simple parameter sharing (poor user differentiation). Our multi-view autoencoder treats each user as a distinct "view," enabling structured interference management through learnable projections with shared backbones, requiring only 2.4\% parameter overhead for 64 users at $\Bar{\rho}=1/12$ and 9.4\% at $\Bar{\rho}=1/6$.}

Many distributed compression and multi-user DeepJSCC systems, such as those described in \cite{mital2022neural,yilmaz2024distributed,wu2023fusion,li2023non,bo2024deep,zhang2024interference,saidutta2020vae}, use distinct encoders and decoders for each user. A practical system, however, should employ parameter sharing to enhance efficiency and scalability. However, challenges arise when all users employ the same mapping function, as users create direct interference and the model struggles to differentiate between users' data.

\modified{Our multi-view framework treats each user as a distinct "view": shared encoder/decoder parameters capture common patterns, learnable user-specific projections $\Pm_i$ create structured separation, joint decoding processes combined signals, and orthogonal initialization ensures training stability.}

We introduce a novel superposition coding method with user-specific projections, which allows network to scale to multiple users without significantly increasing the number of parameters and without changing the DeepJSCC architecture that is known to perform well. The encoders at the users share all their parameters, excluding the user-specific projection layer, i.e., $\Thetav_1=\cdots=\Thetav_n=\Thetav$. At every user $i \in \LSB n \RSB$, we have a complex-valued projection matrix $\Pm \in \CC^{m \times nm}$, where $m$ is the number of filters of the last convolutional layer of the encoder network. Similarly, at the receiver, we have a complex-valued projection matrix $\Pm^{'} \in \CC^{nm \times m}$ for every image to be decoded, i.e., $i \in \LSB n \RSB$. We initialize $\Pm_i$ and $\Pm_i^{'}$, $\forall i \in \LSB n \RSB$, in a way that all of their rows are orthogonal, via QR factorization as detailed in \cref{alg:curriculum}. We jointly optimize projection matrices along with the other parameters of the network throughout the training.

\begin{remark}[CDMA Analogy]
\label{remark:cdma}
These matrices can be considered as the spreading codes in a CDMA system. The list of possible projection matrices can be agreed upon a priori, and these matrices can be shared across all the terminals. Then, at each instance, the active users can be assigned one of these matrices by the receiver, e.g., base station.

\modified{Unlike traditional CDMA spreading codes ($<$1KB per user)~\cite{verdu1998multiuser}, our approach requires projection matrices $\Pm_i \in \mathbb{C}^{m \times nm}$ and $\Pm_i' \in \mathbb{C}^{nm \times m}$ (~32MB per user for $m = 256$, $n = 64$). This overhead enables end-to-end optimization and superior interference management, with matrices distributed once during deployment.}
\end{remark}
Data processing begins with the encoder DNN, which converts an image into real-valued tensor of shape $\RR^{W' \times H' \times 2m}$, where $W'$ and $H'$ are the downsampled dimensions. These tensors are then mapped to complex numbers by pairing consecutive real values, resulting in $\CC^{W' \times H' \times m}$. Each user subsequently applies a user-specific projection by multiplying with $\Pm_i$, resulting in $\Tilde{\zv}_i \Pm_i$, where $\Tilde{\zv}_i$ is the encoder output. The output is flattened and undergoes power normalization to adhere to the average power constraint. This involves scaling the raw signal $\Tilde{\zv}_i$ by $\sqrt{k \Pavg/(\Tilde{\zv}_i \Tilde{\zv}_i^H)}$ after flattening, where $\Tilde{\zv}_i^H$ is the complex conjugate transpose of $\Tilde{\zv}_i$. Finally, each user $i \in \LSB n \RSB$ sends its signal over the channel.

The receiver obtains superposed signals along with multiplicative and additive noise. The receiver first reshapes the matrix to image form of shape $\CC^{W^{'} \times H^{'} \times m}$, and then multiplies the reshaped tensor with $\Pm_i^{'}$, i.e., $\yv \Pm_i^{'}$, which is then given as an input to the decoder DNN to obtain $D_{\Phi} (\yv \Pm_i^{'})$. This projection and decoding step is repeated for every user $i \in \LSB n \RSB$. \modified{This enables parameter sharing while distinguishing user inputs/outputs, improving training and scalability. The multi-view design provides computational efficiency, progressive training compatibility, variable user support, and joint optimization benefits.}
\begin{remark}[Scalability]
\label{remark:scalability}
The overall number of network parameters is bounded by
\[
\Oc\Bigl( |\Thetav| + |\Phiv| + n^2 \Bar{\rho}^2 \Bigr),
\]
where \( |\Thetav| \) and \( |\Phiv| \) denote the sizes of the shared encoder and decoder, respectively, and the term \( n^2 \Bar{\rho}^2 \) accounts for the user-specific projection matrices. Since the dominant contribution comes from the shared parameters \( |\Thetav| + |\Phiv| \), the architecture remains scalable even as the number of users \( n \) grows large.
\end{remark}
\begin{remark}[Shape Adaptivity]
    The network can process input images with arbitrary width $W$ and height $H$ since only the filter dimension is affected by the projection.
\end{remark}
\begin{remark}[Parallelization]
The receiver can decode images from different users in parallel because the decoders are identical, and the projections are independent, requiring no communication. The receiver only needs to assign a particular index to each user so that they know which projection matrix to use. This can be done during the user admission process.
\end{remark}

\subsection{Training Procedure and Construction of Training Samples}
\label{sec:subsampling}

We jointly train the whole network with parameters $\Thetav, \Phiv, \Pm_i, \Pm_i^{'}, \forall i \in \LSB n \RSB$, using the training data samples $\Dctrain$ via the loss function $
    \Lc = \sum_{\substack{(\xv_1, \ldots, \xv_n) \\ \in \Dctrain}} \sum_{i =1}^n \mathrm{MSE} \LB \xv_i, \hat{\xv}_i \RB$,
where $\mathrm{MSE}(\xv,\hat{\xv}) \triangleq \frac{1}{m} \norm{\xv - \hat{\xv}}_2^2$ and $m$ is the total number of elements in $\xv$, which is given by $m=C_\mathrm{in} W H$. Notice that minimizing $\Lc$ also minimizes $\mathrm{PSNR}$ over the training set. We note that the introduced method is unsupervised as it does not rely on any costly human labeling, and raw images and the channel model are enough to train our method. \Cref{alg:overview} shows the training and fine-tuning procedures of DeepJSCC-PNOMA.

\tikzexternaldisable
\begin{algorithm}[t!]
\caption{\strut Training and Fine-tuning of DeepJSCC-PNOMA}
\label{alg:overview}
\begin{algorithmic}
\State Construct $\Dctrain$ and $\Dcval$ for $n$ users
\Comment{See Section~\ref{sec:subsampling}}
\State Initialize $\Thetav$ and $\Phiv$ randomly for $n=1$ or from previous values for $n>1$
\Comment{See Section~\ref{sec:curriculum_learning}}
\State Initialize $\Pm_i$ and $\Pm^{'}_i$, $\forall i \in \LSB n \RSB$, as $I$ for $n=1$ or orthogonally for $n > 1$
\Comment{See Section~\ref{sec:projection}}
\Repeat
\Comment{Iterate through epochs}
    \State Shuffle $\Dctrain$ and set $\Lc=t=0$
    \ForAll{$(\xv_1,\ldots) \in \Dctrain$}
        \For{$i\in \LSB n\RSB$}
        \Comment{Device with index $i$}
            \State $\Tilde{\zv}_i = E_{\Thetav} (\xv_i,h_i,\sigma)$
            \Comment{Encoding}
            \State $\zv_i = \mathrm{Flatten}(\Tilde{\zv}_i \Pm_i$ )
            \Comment{User-specific projection}
            \State $\zv_i = \sqrt{k \Pavg/(\zv_i \zv_i^H)} \zv_i$
            \Comment{Power normalization}
        \EndFor
        \LComment{Transmission of latents over a \gls{MAC}}
        \State Calculate $\sigma$ via~\eqref{eq:snr} for $\mathrm{SNR} \sim \mathrm{Uniform}\LSB 0,20 \RSB$
        \State Sample $\nv \sim \CC\Nc(\vec{0}, \sigma^2 \vec{I}_{k})$
        \If{channel is Gaussian}
            \State $h_i=1, \, \forall i \in \LSB n \RSB$
        \ElsIf{channel is Rayleigh}
            \State Sample $h_i \sim \CC\Nc(0, 1), \, \forall i \in \LSB n \RSB$
        \EndIf
        \State $\yv = \sum_{i=1}^n h_i \zv_i + \nv$
        \Comment{Received signal}
        \For{$i\in \LSB n\RSB$}
            \State $\yv_i = \mathrm{Reshape}(\Pm_i^{'} \yv)$
            \Comment{User-specific projection}
            \State $\hat{\xv}_i = D_{\Phi} \LB \yv_i,h_i,\sigma \RB$
            \Comment{Decoding}
        \EndFor
        \State $\Lc = \Lc +  \sum_{i =1}^n \mathrm{MSE} \LB \xv_i, \hat{\xv}_i \RB$
        \State t = t + 1
        \If{$t \ \mathrm{mod} \  batch \  size = 0$}
            \LComment{Standard mini-batch training}
            \State Update $\Thetav, \Phiv, \Pm_i, \Pm^{'}_i, \  \forall i \in \LSB n \RSB$ by backpropagation 
            \State Reset $\Lc$ and gradients to zero
        \EndIf
    \EndFor
    \State Compute validation PSNR using $\Dcval$
\Until{validation PSNR improves less than $\Delta$ for $e$ epochs}
\end{algorithmic}
\end{algorithm}
\tikzexternalenable

During training, our neural network takes multiple image instances $\xv_1,\ldots,\xv_n$ as input and decodes these images as $\hat{\xv}_1, \ldots, \hat{\xv}_n$. \modified{Therefore, our task aligns well with the \gls{MTL} framework~\cite{zhang2021survey}, as reconstructing the image of each user can be viewed as a distinct yet closely related task.} One common method in \gls{MTL} is to share parameters to model commonalities between tasks. \modified{In our problem, since the task of image transmission for all encoders is exactly equivalent, we naturally share parameters among the encoders.} However, the receiver also needs to distinguish between the images transmitted by different encoders and associate each decoded image to its transmitter during the decoding phase. This is why we employ a user-specific projection layer, which is explained in the next section. The projection matrices $\Pm_i$ and $\Pm^{'}_i$, $\forall i \in \LSB n \RSB$, are optimized jointly with the rest of the network during training and remain constant during test time. \modified{This simple method allows the use of a standard single-user DeepJSCC architecture, spreading its output over the available bandwidth while obeying the average power constraint and enabling the decoder to differentiate between the transmitters of different images.}

\modified{We now define the training tuples in $\Dctrain$ to compute the loss function $\Lc$.} Since we are training the network on $(\xv_1, \xv_2, \ldots,\xv_n)$ tuples, there are $N^n$ possible combinations for a given training dataset with $N$ samples, i.e., $\card{\Dctrain}=N$. However, it is generally infeasible to use all of these combinations since the number of training instances grows exponentially with $N$. Moreover, our experimental results suggest that distributing random samples from a shuffled training batch to different users may cause over-regularization and lead to underfitting. Consistently using the same sample combinations from a finite list of tuples across different epochs improved the outcomes.

Therefore, we subsample $T \ll N^n$ tuples from all possible combinations~\cite{hoffer2015deep}. \modified{To generate $T$ tuples, we sample $nT$ integers from $\LSB N \RSB$ and group every $n$ consecutive samples.} We use the same tuples in $\Dctrain$ after shuffling to minimize $\Lc$. \modified{We also construct validation tuples $\Dcval$ by splitting the initial validation data among $n$ users and use the same validation tuples after every epoch.} The algorithms in \appendixConstruction \footnote{Due to page constraints, the appendix is omitted here but is fully available on arXiv at \href{https://arxiv.org/abs/2504.03690}{arXiv:2504.03690}.}\ present the methodology for constructing training and evaluation samples, respectively. %

\subsection{Novel Progressive Fine-Tuning Method for Doubling the Number of Users}
\label{sec:curriculum_learning}

\newcommand{\progressiveTrainingDiagramCaption}{\modified{Progressive fine-tuning methodology for scaling DeepJSCC-PNOMA from 1 to 16 users. The process iteratively doubles the number of users while maintaining training stability through orthogonal initialization. Each stage preserves signal power and ensures initial orthogonality between user groups, enabling convergence from interference-free configurations.}}

\newcommand{\progressiveTrainingDiagramContent}{%
\begin{tikzpicture}[
    node distance=2cm and 4cm,
    box/.style={rectangle, draw, fill=blue!10, text width=2.2cm, text centered, minimum height=1.2cm, rounded corners=2pt},
    first_box/.style={rectangle, draw, fill=green!10, text width=3.0cm, text centered, minimum height=1.2cm, rounded corners=2pt},
    math_box/.style={rectangle, draw, fill=yellow!20, text width=3.2cm, text centered, minimum height=1.8cm, rounded corners=2pt},
    process/.style={rectangle, draw, fill=green!10, text width=2.8cm, text centered, minimum height=1.2cm, rounded corners=2pt},
    arrow/.style={->, very thick, blue!70, line width=1.2pt, >=stealth},
    stage_num/.style={circle, draw=blue!70, fill=blue!20, inner sep=2pt, font=\footnotesize\bfseries, minimum size=12pt, text=blue!80},
    stage_label/.style={font=\small\bfseries, text=blue!80, draw=blue!30, fill=blue!5, rounded corners=3pt, inner sep=3pt},
    small_text_box/.style={rectangle, draw=blue!50, fill=blue!5, text width=1.5cm, text centered, minimum height=0.6cm, rounded corners=2pt, font=\tiny}
]

\node[stage_label] at (0, 4.5) {Stage 1};
\node[stage_label] at (4.0, 4.5) {Stage 2};
\node[stage_label] at (8.0, 4.5) {Stage 3};
\node[stage_label] at (12.0, 4.5) {Stage 4};

\node[first_box] (train1) at (0, 3.5) {Train DeepJSCC\\1 user\\$\Pm = \Id_m$};
\node[box] (train2) at (4.0, 3.5) {Fine-tune\\2 users\\$n=2$};
\node[box] (train3) at (8.0, 3.5) {Fine-tune\\4 users\\$n=4$};
\node[box] (train4) at (12.0, 3.5) {Fine-tune\\8 users\\$n=8$};

\node[math_box] (math_expand) at (2.0, 1.0) {$\Pm \leftarrow \begin{bmatrix} \Pm \Km_1 \\ \Pm \Km_2 \end{bmatrix}$\\[0.2em]\rule{2.5cm}{0.4pt}\\[0.2em]$\Pm' \leftarrow \begin{bmatrix} \Km_1^H \Pm' \\ \Km_2^H \Pm' \end{bmatrix}$};
\node[math_box] (math2) at (6.0, 1.0) {$\Pm \leftarrow \begin{bmatrix} \Pm \Km_1 \\ \Pm \Km_2 \end{bmatrix}$\\[0.2em]\rule{2.5cm}{0.4pt}\\[0.2em]$\Pm' \leftarrow \begin{bmatrix} \Km_1^H \Pm' \\ \Km_2^H \Pm' \end{bmatrix}$};
\node[math_box] (math3) at (10.0, 1.0) {$\Pm \leftarrow \begin{bmatrix} \Pm \Km_1 \\ \Pm \Km_2 \end{bmatrix}$\\[0.2em]\rule{2.5cm}{0.4pt}\\[0.2em]$\Pm' \leftarrow \begin{bmatrix} \Km_1^H \Pm' \\ \Km_2^H \Pm' \end{bmatrix}$};
\node (dots) at (13.0, 1.3) {\large $\cdots$};
\draw[arrow] (train4) -- ++(1.0, -1.75);


\draw[arrow] (train1) -- (math_expand);
\draw[arrow] (math_expand) -- (train2);
\draw[arrow] (train2) -- (math2);
\draw[arrow] (math2) -- (train3);
\draw[arrow] (train3) -- (math3);
\draw[arrow] (math3) -- (train4);

\node[small_text_box] at (2.0, 2.45) {Generate orthogonal\\matrices $\Km_1, \Km_2$};
\node[small_text_box] at (6.0, 2.45) {Generate orthogonal\\matrices $\Km_1, \Km_2$};
\node[small_text_box] at (10.0, 2.45) {Generate orthogonal\\matrices $\Km_1, \Km_2$};



\end{tikzpicture}%
}

\ifdefined\tworeviewresponse
\centering
\captionof{figure}{\progressiveTrainingDiagramCaption}
\label{fig:progressive_training}
\progressiveTrainingDiagramContent
\else
\begin{figure*}[t]
\centering
\caption{\progressiveTrainingDiagramCaption}
\label{fig:progressive_training}
\progressiveTrainingDiagramContent
\end{figure*}
\fi

It is known that high noise in the data, gradients or weights adversely affect the training of \glspl{DNN}. In a \gls{MAC}, we not only deal with additive and multiplicative noise effects but also interference from other users since signals are superposed.

When $\Pm_i$ and $\Pm^{'}_i$, $\forall i \in \LSB n \RSB$, are initialized randomly, signals of different devices interfere with each other, and the amount of interference increases with the number of users. We address this problem via a novel progressive fine-tuning method by gradually doubling the number of users while sustaining the performance of fine-tuned network in the beginning of training via orthogonal initialization of projection matrices. 

\modified{\Cref{fig:progressive_training} illustrates our progressive fine-tuning methodology, which enables stable scaling from single-user to multi-user configurations. The detailed algorithmic procedure is presented in \Cref{alg:curriculum}.}
\tikzexternaldisable
\begin{algorithm}[t!]
\caption{\strut Progressive fine-tuning for doubling the number of users in DeepJSCC-PNOMA}
\label{alg:curriculum}
\begin{algorithmic}
    \State Initialize $\Thetav$ and $\Phiv$ randomly for a given $\Bar{\rho}$
    \State $k=3WH\Bar{\rho}=3WH\rho$ \Comment{k: available bandwidth for \gls{TDMA} case ($n=1$) given $\Bar{\rho}$}.
    \State $m=\frac{k}{WH/4^c} = \frac{3WH\rho}{WH/4^c} = 3\rho 4^c$ \Comment{m: \# of transmitted filters}
    \State $\Pm_1 = \Pm_1^{'} = \mathbf{I}_m$
    \LComment{Initial Training}
    \State Train $\Thetav$ and $\Phiv$ using \cref{alg:overview}
    \State $\Pm_{1,\mathrm{old}} = \Pm_1$
    \State $\Pm^{'}_{1,\mathrm{old}} = \Pm^{'}_1$
    \ForAll{$n \in \LB 2^1, 2^2, 2^3, \ldots \RB$} 
        \LComment{Generate orthogonal matrix $\Qm$}
        \State Sample $\Mm \sim \Nc \LB \mathbf{0}_{nm}, \frac{1}{nm} \mathbf{I}_{nm} \RB$
        \State $\Qm, \Rm = \mathrm{QRFactorization} \LB \Mm \RB$
        \LComment{Make $\Qm$ uniform according to the procedure in~\cite{mezzadri2006generate}}
        \State $\dv = \mathrm{Sign} \LB \mathrm{Diagonal} \LB \Rm \RB \RB$
        \State $\begin{bmatrix}
            \Km_1 \\ \Km_2
        \end{bmatrix} = \dv \cdot \Qm$ \Comment{$\Km_1, \Km_2 \in \CC^{\frac{n}{2}m \times nm}$}
        \ForAll{$i \in \LB 1,2,\ldots,\nicefrac{n}{2} \RB$}
            \LComment{Copy and orthogonalize projection matrices of first half of devices}
            \State $\Pm_i = \Pm_{i,old} \Km_{1}$
            \State $\Pm_i^{'} = \Km_{1}^H \Pm_{i,old}^{'}$
            \LComment{Copy and orthogonalize projection matrices of second half of devices}
            \State $\Pm_{i+\nicefrac{n}{2}} = \Pm_{i,old} \Km_{2}$
            \State $\Pm_{i+\nicefrac{n}{2}}^{'} = \Km_{2}^{H} \Pm_{i,old}^{'}$
        \EndFor
        \LComment{Fine-tuning step}
        \State Fine-tune $\Thetav$, $\Phiv$, $\Pm$ and $\Pm^{'}_i, \, \forall i \in \LSB n \RSB$ using \cref{alg:overview}
        \LComment{Copy projection matrices for the next step}
        \ForAll{$i \in \LSB n \RSB$}
            \State $\Pm_{i,\mathrm{old}} = \Pm_i$
            \State $\Pm^{'}_{i,\mathrm{old}} = \Pm^{'}_i$
        \EndFor
        \State Export DeepJSCC-PNOMA with $n$ users: $E_{\Thetav}$, $D_{\Phiv}$, $\Thetav$, $\Phiv$, $\Pm_i$ and $\Pm^{'}_i, \, \forall i \in \LSB n \RSB$
        \EndFor
    \end{algorithmic}
\end{algorithm}
\tikzexternalenable

We first start with training a standard point-to-point DeepJSCC (or DeepJSCC-TDMA), i.e., DeepJSCC-PNOMA when $n=1$ and $\mathbf{P}_1=\mathbf{I}_m$, where $m$ is the number of filters in the transmitted signal that is processed by \glspl{CNN}. Given a trained or fine-tuned network for $\nicefrac{n}{2}$ users, i.e. $\Thetav$, $\Phiv$ and $\Pm_i$, $\Pm_i^{'}$, $\forall i \in \LSB \nicefrac{n}{2} \RSB$, we now describe how we extend it to $n$ users as follows. We first create a new encoder $E$ and decoder $D$ for each user using the previous parameters $\Thetav$ and $\Phiv$ with parameter sharing, respectively. We then copy the previous projections to $\Pm_{i,\mathrm{old}} = \Pm_i$ and $\Pm_{i,\mathrm{old}}^{'} = \Pm_i^{'}$, $\forall i \in \LSB \nicefrac{n}{2} \RSB$. Then, we duplicate all these projection matrices so that $\Pm_i=\Pm_{\nicefrac{n}{2}+i}=\Pm_{i,\mathrm{old}}$ and $\Pm_i^{'}=\Pm_{\nicefrac{n}{2}+i}^{'}=\Pm_{i,\mathrm{old}}^{'}$. We then generate a uniformly distributed orthonormal matrix by sampling an \(nm \times nm\) Gaussian matrix $\mathbf{M} \sim \mathcal{N}\left(\mathbf{0}, \frac{1}{nm}\mathbf{I}_{nm}\right)$ and computing its QR factorization, $\mathbf{M} = \mathbf{Q}\mathbf{R}$. To remove the sign (or phase) ambiguity in the QR decomposition, we post-multiply \(\mathbf{Q}\) by a diagonal matrix with entries~\cite{mezzadri2006generate}
\[
\mathbf{d} = \operatorname{Sign}\bigl(\operatorname{Diagonal}(\mathbf{R})\bigr),
\]
and partition the adjusted matrix as
\[
\begin{bmatrix}
\Km_1 \\[1mm]
\Km_2
\end{bmatrix} = \mathbf{d} \cdot \mathbf{Q},
\]
where \(\Km_1, \Km_2 \in \mathbb{C}^{\frac{n}{2}m \times nm}\).

We multiply $\Pm_i$, $\forall i \in \LSB \nicefrac{n}{2} \RSB$, with $\Km_1$ and multiply $\Pm_i$, $\forall i \in \LSB \nicefrac{n}{2}+1, n \RSB$, with $\Km_2$. We also multiply $\Pm_i^{'}$, $\forall i \in \LSB \nicefrac{n}{2} \RSB$, with $\Km_1^{H}$ and multiply $\Pm_i^{'}$, $\forall i \in \LSB \nicefrac{n}{2}+1, n \RSB$, with $\Km_2^{H}$. Notice that it is enough to optimize $\Pm_i$ and $\Pm_i^{'}$ instead of their factors due to the linearity property of matrix multiplication, allowing efficiency in optimization.

At this stage, it is crucial to note that the power of the encoded signal remains unchanged after applying the projection matrices. \modified{In other words, for each \(i \in \{1, \dots, n\}\), the encoded signal \(\tilde{\zv}_i\) satisfies
\[
\|\tilde{\zv}_i\|_2^2 = \|\tilde{\zv}_i \Pm_i\|_2^2.
\]
This property follows from the fact that each projection matrix \(\Pm_i\) is unitary (i.e., \(\Pm_i^H \Pm_i = I\)); specifically, since \(\|\tilde{\zv}_i\|_2^2 = \tilde{\zv}_i^H \tilde{\zv}_i\) and 
\[
\|\tilde{\zv}_i \Pm_i\|_2^2 = (\tilde{\zv}_i \Pm_i)^H (\tilde{\zv}_i \Pm_i) = \tilde{\zv}_i^H (\Pm_i^H \Pm_i) \tilde{\zv}_i = \tilde{\zv}_i^H \tilde{\zv}_i,
\]
the power is preserved.}

\modified{Moreover, due to the orthogonality of the new projection matrices \(\Km_1\) and \(\Km_2\) (i.e., \(\Km_1 \Km_2^H = 0\)), the normalized signals corresponding to different user groups are guaranteed to be orthogonal at this stage before any optimization. In particular, for a user \(i\) (using \(\Km_1\)) and a user \(j\) (using \(\Km_2\)), the inner product of their normalized signals is
\[
\zv_i \cdot \zv_j^H = \frac{\tilde{\zv}_i \Pm_i \Km_1}{\sqrt{\frac{1}{k}\|\tilde{\zv}_i \Pm_i \Km_1\|^2}} \cdot \left(\frac{\tilde{\zv}_j \Pm_j \Km_2}{\sqrt{\frac{1}{k}\|\tilde{\zv}_j \Pm_j \Km_2\|^2}}\right)^H = 0,
\]
which follows immediately from \(\Km_1 \Km_2^H = 0\).}

Due to the design of the user-specific projection matrices in DeepJSCC-PNOMA—which preserve the power of each encoded signal and enforce mutual orthogonality among signals—the system with \(n\) users initially behaves as if it were time-sharing among \(\frac{n}{2}\) users under fixed per-user average power \(\Bar{P}_\mathrm{avg}\) and per-user bandwidth ratio \(\Bar{\rho}\). In this configuration, each user’s signal retains its original power and remains free from interference, enabling independent decoding. Consequently, the network can begin fine-tuning without any performance degradation from inter-user interference.
\changed{By gradually increasing the system complexity while maintaining stability and efficiency, this progressive training approach serves as an effective curriculum for multi-user learning. The encoder–decoder pair is guaranteed to learn a robust feature representation without any interference from other users when they begin with a single-user DeepJSCC baseline. In order to guarantee that each user initially occupies a unique and interference-free subspace in the transmission domain, previously learned projections are replicated and orthogonalized using QR-based transformations when the number of users doubles. The power of every encoded signal stays constant throughout all stages due to the unitary nature of the projection matrices. By fine-tuning at every stage of expansion, the network can seamlessly adjust to the new user configuration and gradually learn to handle lingering interference and user correlation. As a result, the model converges faster, avoids instability associated with joint multi-user initialization, and achieves improved generalization through structured and incremental adaptation.}
The next step is to fine-tune the model using the same training procedure described in \cref{sec:subsampling}.
\section{Numerical Results}
\label{sec:numerical_results}
\newcommand{\separationawgnfigureCIFAR}[6]{
\centering
\begin{tikzpicture}[scale=0.5]
        \begin{axis}[
        title={#1},
        xlabel={$\mathrm{SNR}_\mathrm{test}$ (\si{\decibel})},
        error bars/y dir=both,
        error bars/y explicit,
        ylabel={#4},
        cycle list/Set1-6,
        cycle multiindex* list={
                [6 of]mark list*\nextlist
                Set1-6\nextlist                {solid,solid,solid,solid,solid,solid,dashed,dashed,dashed,dashed,dashed,dashed}\nextlist
        },
        xmin=0,
        xmax=15,
        tick label style={/pgf/number format/fixed}
        ]
        \foreach \bwfactor in {6,12}{
            \addplot table[x=model/snr, y=test/#5, y error=test/#5_std, col sep=comma]{results/#2/#3_rho\bwfactor_noma_nd16.csv};\label{separationawgnfigure_#2_#3_#5_rho\bwfactor_noma_nd16_#6}
            \addplot table[x=model/snr, y=test/#5, y error=test/#5_std, col sep=comma]{results/#2/#3_rho\bwfactor_tdma.csv};\label{separationawgnfigure_#2_#3_#5_rho\bwfactor_tdma_#6}
            \addplot table[x=model/snr, y=test/#5, y error=test/#5_std, col sep=comma]{results/separation/#2/#3_rho\bwfactor_bpg_#6.csv};\label{separationawgnfigure_#2_#3_#5_rho\bwfactor_bpg_#6}
            \addplot[draw=none] coordinates {(1,1)};
            \addplot[draw=none] coordinates {(1,1)};
            \addplot table[x=model/snr, y=test/#5, y error=test/#5_std, col sep=comma]{results/separation/#2/#3_rho\bwfactor_bmshj2018_#6.csv};\label{separationawgnfigure_#2_#3_#5_rho\bwfactor_bmshj2018_#6}
        }
        \end{axis}
        \end{tikzpicture}
}
\newcommand{\separationawgnfigureCIFARbig}[6]{
\centering
\begin{tikzpicture}[scale=0.65]
        \begin{axis}[
        title={#1},
        xlabel={$\mathrm{SNR}_\mathrm{test}$ (\si{\decibel})},
        error bars/y dir=both,
        error bars/y explicit,
        ylabel={#4},
        cycle list/Set1-6,
        cycle multiindex* list={
                [6 of]mark list*\nextlist
                Set1-6\nextlist                {solid,solid,solid,solid,solid,solid,dashed,dashed,dashed,dashed,dashed,dashed}\nextlist
        },
        xmin=0,
        xmax=15,
        tick label style={/pgf/number format/fixed}
        ]
        \foreach \bwfactor in {6,12}{
            \addplot table[x=model/snr, y=test/#5, y error=test/#5_std, col sep=comma]{results/#2/#3_rho\bwfactor_noma_nd16.csv};\label{separationawgnfigure_#2_#3_#5_rho\bwfactor_noma_nd16_#6}
            \addplot table[x=model/snr, y=test/#5, y error=test/#5_std, col sep=comma]{results/#2/#3_rho\bwfactor_tdma.csv};\label{separationawgnfigure_#2_#3_#5_rho\bwfactor_tdma_#6}
            \addplot table[x=model/snr, y=test/#5, y error=test/#5_std, col sep=comma]{results/separation/#2/#3_rho\bwfactor_bpg_#6.csv};\label{separationawgnfigure_#2_#3_#5_rho\bwfactor_bpg_#6}
            \addplot[draw=none] coordinates {(1,1)};
            \addplot[draw=none] coordinates {(1,1)};
            \addplot table[x=model/snr, y=test/#5, y error=test/#5_std, col sep=comma]{results/separation/#2/#3_rho\bwfactor_bmshj2018_#6.csv};\label{separationawgnfigure_#2_#3_#5_rho\bwfactor_bmshj2018_#6}
        }
        \end{axis}
        \end{tikzpicture}
}

\newcommand{\separationawgnfigure}[6]{
        \centering
\begin{tikzpicture}[scale=0.5]
        \begin{axis}[
        title={#1},
        xlabel={$\mathrm{SNR}_\mathrm{test}$ (\si{\decibel})},
        error bars/y dir=both,
        error bars/y explicit,
        ylabel={#4},
        cycle list/Set1-6,
        cycle multiindex* list={
                [6 of]mark list*\nextlist
                Set1-6\nextlist                {solid,solid,solid,solid,solid,solid,dashed,dashed,dashed,dashed,dashed,dashed}\nextlist
        },
        xmin=0,
        xmax=15,
        tick label style={/pgf/number format/fixed}
        ]
        \foreach \bwfactor in {6,12}{
            \addplot table[x=model/snr, y=test/#5, y error=test/#5_std, col sep=comma]{results/#2/#3_rho\bwfactor_noma_nd16.csv};\label{separationawgnfigure_#2_#3_#5_rho\bwfactor_noma_nd16_#6}
            \addplot table[x=model/snr, y=test/#5, y error=test/#5_std, col sep=comma]{results/#2/#3_rho\bwfactor_tdma.csv};\label{separationawgnfigure_#2_#3_#5_rho\bwfactor_tdma_#6}
            \addplot table[x=model/snr, y=test/#5, y error=test/#5_std, col sep=comma]{results/separation/#2/#3_rho\bwfactor_bpg_#6.csv};\label{separationawgnfigure_#2_#3_#5_rho\bwfactor_bpg_#6}
            \addplot table[x=model/snr, y=test/#5, y error=test/#5_std, col sep=comma]{results/separation/#2/#3_rho\bwfactor_mlicpp_#6.csv};\label{separationawgnfigure_#2_#3_#5_rho\bwfactor_mlicpp_#6}
            \addplot table[x=model/snr, y=test/#5, y error=test/#5_std, col sep=comma]{results/separation/#2/#3_rho\bwfactor_cheng2020_anchor_#6.csv};\label{separationawgnfigure_#2_#3_#5_rho\bwfactor_cheng2020_anchor_#6}
            \addplot table[x=model/snr, y=test/#5, y error=test/#5_std, col sep=comma]{results/separation/#2/#3_rho\bwfactor_bmshj2018_#6.csv};\label{separationawgnfigure_#2_#3_#5_rho\bwfactor_bmshj2018_#6}
        }
        \end{axis}
        \end{tikzpicture}
}

\newcommand{\separationawgnfigurebig}[6]{
        \centering
\begin{tikzpicture}[scale=0.65]
        \begin{axis}[
        title={#1},
        xlabel={$\mathrm{SNR}_\mathrm{test}$ (\si{\decibel})},
        error bars/y dir=both,
        error bars/y explicit,
        ylabel={#4},
        cycle list/Set1-6,
        cycle multiindex* list={
                [6 of]mark list*\nextlist
                Set1-6\nextlist                {solid,solid,solid,solid,solid,solid,dashed,dashed,dashed,dashed,dashed,dashed}\nextlist
        },
        xmin=0,
        xmax=15,
        tick label style={/pgf/number format/fixed}
        ]
        \foreach \bwfactor in {6,12}{
            \addplot table[x=model/snr, y=test/#5, y error=test/#5_std, col sep=comma]{results/#2/#3_rho\bwfactor_noma_nd16.csv};\label{separationawgnfigure_#2_#3_#5_rho\bwfactor_noma_nd16_#6}
            \addplot table[x=model/snr, y=test/#5, y error=test/#5_std, col sep=comma]{results/#2/#3_rho\bwfactor_tdma.csv};\label{separationawgnfigure_#2_#3_#5_rho\bwfactor_tdma_#6}
            \addplot table[x=model/snr, y=test/#5, y error=test/#5_std, col sep=comma]{results/separation/#2/#3_rho\bwfactor_bpg_#6.csv};\label{separationawgnfigure_#2_#3_#5_rho\bwfactor_bpg_#6}
            \addplot table[x=model/snr, y=test/#5, y error=test/#5_std, col sep=comma]{results/separation/#2/#3_rho\bwfactor_mlicpp_#6.csv};\label{separationawgnfigure_#2_#3_#5_rho\bwfactor_mlicpp_#6}
            \addplot table[x=model/snr, y=test/#5, y error=test/#5_std, col sep=comma]{results/separation/#2/#3_rho\bwfactor_cheng2020_anchor_#6.csv};\label{separationawgnfigure_#2_#3_#5_rho\bwfactor_cheng2020_anchor_#6}
            \addplot table[x=model/snr, y=test/#5, y error=test/#5_std, col sep=comma]{results/separation/#2/#3_rho\bwfactor_bmshj2018_#6.csv};\label{separationawgnfigure_#2_#3_#5_rho\bwfactor_bmshj2018_#6}
        }
        \end{axis}
        \end{tikzpicture}
}

\begin{figure*}
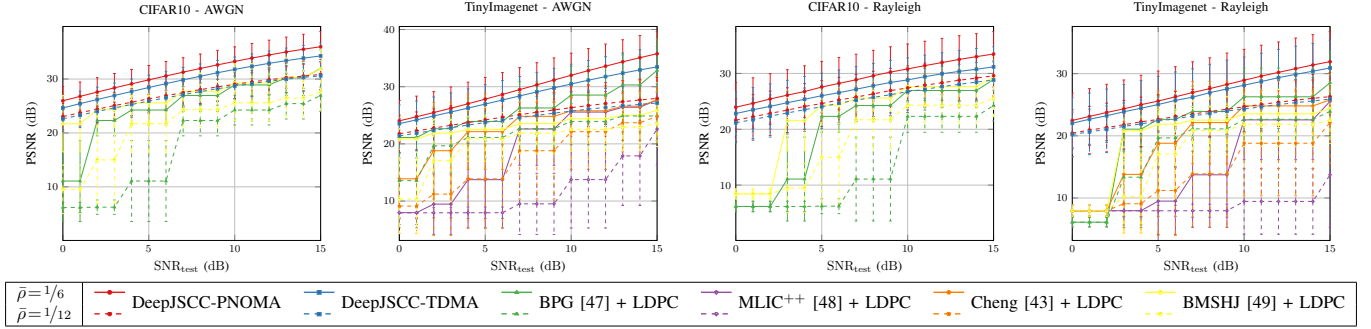

\separationawgnfigureCIFAR{CIFAR10 - AWGN}{CIFAR10}{awgn}{PSNR (\si{dB})}{psnr}{ldpc}
\separationawgnfigure{TinyImagenet - AWGN}{TinyImagenet}{awgn}{PSNR (\si{dB})}{psnr}{ldpc}
\separationawgnfigureCIFAR{CIFAR10 - Rayleigh}{CIFAR10}{rayleigh}{PSNR (\si{dB})}{psnr}{ldpc}
\separationawgnfigure{TinyImagenet - Rayleigh}{TinyImagenet}{rayleigh}{PSNR (\si{dB})}{psnr}{ldpc}
\resizebox{\textwidth}{!}{
\begin{tabular}{|l|llllll|}
\hline
$\Bar{\rho}=\nicefrac{1}{6}$ &
  \ref*{separationawgnfigure_TinyImagenet_awgn_psnr_rho6_noma_nd16_ldpc}
  \multirow{2}{*}{DeepJSCC-PNOMA} &
  \ref*{separationawgnfigure_TinyImagenet_awgn_psnr_rho6_tdma_ldpc}
  \multirow{2}{*}{DeepJSCC-TDMA} &
  \ref*{separationawgnfigure_TinyImagenet_awgn_psnr_rho6_bpg_ldpc}
  \multirow{2}{*}{BPG~\cite{Bellard:BPG} + LDPC} &
  \ref*{separationawgnfigure_TinyImagenet_awgn_psnr_rho6_mlicpp_ldpc}
  \multirow{2}{*}{MLIC$^{++}$~\cite{jiang2023mlicpp} + LDPC} &
  \ref*{separationawgnfigure_TinyImagenet_awgn_psnr_rho6_cheng2020_anchor_ldpc}
  \multirow{2}{*}{Cheng~\cite{cheng2020learned} + LDPC} &
  \ref*{separationawgnfigure_TinyImagenet_awgn_psnr_rho6_bmshj2018_ldpc}
  \multirow{2}{*}{BMSHJ~\cite{ballé2018variational} + LDPC}  \\
$\Bar{\rho}=\nicefrac{1}{12}$ &
  \ref*{separationawgnfigure_TinyImagenet_awgn_psnr_rho12_noma_nd16_ldpc} &
  \ref*{separationawgnfigure_TinyImagenet_awgn_psnr_rho12_tdma_ldpc} &
  \ref*{separationawgnfigure_TinyImagenet_awgn_psnr_rho12_bpg_ldpc} &
  \ref*{separationawgnfigure_TinyImagenet_awgn_psnr_rho12_mlicpp_ldpc} &
  \ref*{separationawgnfigure_TinyImagenet_awgn_psnr_rho12_cheng2020_anchor_ldpc} &
  \ref*{separationawgnfigure_TinyImagenet_awgn_psnr_rho12_bmshj2018_ldpc}
   \\ \hline
\end{tabular}
}
\caption{Comparison with point-to-point DeepJSCC and separation-based methods.}
\label{fig:comparison_point2point}
\end{figure*}

\subsection{Datasets}
\label{sec:datasets}

We use the CIFAR10 dataset\modified{~\cite{krizhevsky2009learning}} for training and testing. CIFAR10 consists of \num{50000} training images and \num{10000} test images, each of dimensions $3 \times 32 \times 32$. We further split the training set into \num{45000} training instances and \num{5000} validation instances. In addition, we employ the TinyImagenet dataset, which contains \num{100000} training instances, \num{10000} validation instances, and \num{10000} test instances, all with the shape $3 \times 64 \times 64$. For evaluating correlated sources, we use the Cityscapes dataset\modified{~\cite{cordts2016cityscapes}}, which comprises \num{5000} stereo image pairs. Specifically, \num{2975} pairs are allocated for training, while \num{500} and \num{1525} pairs are used for validation and testing, respectively. Following~\cite{yilmaz2024distributed}, each image in the Cityscapes dataset is downsampled to $128 \times 256$. Finally, to assess generalization performance and enable qualitative comparisons, we employ the Kodak dataset, which consists of \num{24} images of dimensions $3 \times 512 \times 768$. Given its limited size, the Kodak dataset is reserved exclusively for testing. Further details on these datasets can be found in \appendixDatasets.

\subsection{Implementation Details}
\label{sec:appendix_implementation}
\begin{table}[t!]
\caption{Employed hyperparameters for fair comparison between scenarios with different number of users}
\label{tab:model_params}
\centering
\begin{tabular}{@{}ccclrrr@{}}
\toprule
$\Bar{\rho}$ & $\Bar{P}_\mathrm{avg}$ & $\sigma^2$ &    Method & n  & $\rho$ & $\Pavg$ \\ \midrule

\multirow{7}{*}{1/6} &  \multirow{7}{*}{1} &  \multirow{7}{*}{1} & DeepJSCC-TDMA & 1 & 1/6 & 1 \\ \cmidrule{4-7}
 & & & \multirow{5}{*}{DeepJSCC-PNOMA} &  1  & 1/6      & 1       \\
                    &   &      &                  & 2  & 2/6    & 1/2     \\
                    &   &       &                 & 4  & 4/6    & 1/4     \\
                   &    &        &                & 8  & 8/6    & 1/8     \\
                    &   &         &               & 16 & 16/6   & 1/16    \\ \cmidrule{4-7}
                    & & & Perfect SIC (2 users) & 1 & 2/6 & 1/2 \\
                    & & & Perfect SIC (16 users) & 1 & 16/6 & 1/16 \\
\midrule
\multirow{7}{*}{1/12} &  \multirow{7}{*}{1} &  \multirow{7}{*}{1} & DeepJSCC-TDMA & 1 & 1/12 & 1 \\ \cmidrule{4-7}
 & & & \multirow{5}{*}{DeepJSCC-PNOMA} &  1  & 1/12      & 1       \\
                    &   &      &                  & 2  & 2/12    & 1/2     \\
                    &   &       &                 & 4  & 4/12    & 1/4     \\
                   &    &        &                & 8  & 8/12    & 1/8     \\
                    &   &         &               & 16 & 16/12   & 1/16    \\ \cmidrule{4-7}
                    & & & Perfect SIC (2 users) & 1 & 2/12 & 1/2 \\
                    & & & Perfect SIC (16 users) & 1 & 16/12 & 1/16 \\
                           \bottomrule
\end{tabular}
\end{table}

We have conducted experiments using the Pytorch framework~\cite{paszke2019pytorch}. We use the same hyperparameters and the same architecture for all the methods. Following previous works for point-to-point DeepJSCC~\cite{bourtsoulatze2019deep,tung2022deepjscc}, we use learning rate \num{1e-4}, set the number of filters in the middle CNN layers to $256$ and batch size to \num{32}. We use Adam optimizer to minimize the loss~\cite{kingma2014adam}. We continue training until no more than $\Delta=1e-3$ improvement is achieved for consecutive $e=10$ epochs. During training and validation, we run the model using different \gls{SNR} values for each instance, uniformly chosen from $\LSB 0,20 \RSB$ \si{\decibel}. We test and report the results for each \gls{SNR} value using the same model. For the proposed method, we use the training data with $3N$ randomly sampled pairs, where elements are chosen among the same training set with $N$ instances, instead of $N^n$ pairs for both CIFAR10 and TinyImagenet datasets, using the method described in \cref{sec:subsampling}. We shuffle the training pairs or instances randomly before each epoch.

\begin{remark}[Comparing Methods with Different Number of Users]
\label{remark:fair_comparison}
    We want to compare scenarios with different number of users fairly. Hence, we fix the per-user bandwidth ratio, $\Bar{\rho}$. This would mean that if we increase the number of users and apply \gls{TDMA}, we would still have the same bandwidth ratio for each image. In order to normalize also the available power, we fix the per-user average power constraint, defined as $\Bar{P}_\mathrm{avg}=\nicefrac{\Pavg}{n}$.
\end{remark}

\Cref{tab:model_params} shows the model hyperparameters used in the experiments for fairness when comparing methods with different numbers of users. DeepJSCC-PNOMA method corresponds to different settings under different hyperparameters as shown in \cref{tab:model_params}, e.g., it can simulate Perfect SIC and DeepJSCC-TDMA under specific hyperparameters.

\subsection{Comparison with Point-to-Point and NOMA-Based Schemes}
\label{sec:comparison_point2point}

\newcommand{\baselines}[6]{
        \centering
\begin{tikzpicture}[scale=0.8]
        \begin{axis}[
        title={#1},
        xlabel={$\mathrm{SNR}_\mathrm{test}$ (\si{\decibel})},
        error bars/y dir=both,
        error bars/y explicit,
        ylabel={#4},
        cycle list/Set1-7,
        cycle multiindex* list={
                [7 of]mark list*\nextlist
                Set1-7\nextlist                {dotted,dotted,solid,solid,dashed,dashed,solid}\nextlist
        },
        xmin=0,
        xmax=15,
        tick label style={/pgf/number format/fixed}
        ]
        \addplot table[x=model/snr, y=test/#5, y error=test/#5_std, col sep=comma]{results/perfectsic/#2_#3_rho#6_perfectsic_nd16.csv};
        \label{baselines_#2_#3_rho#6_perfectsic_nd16}
        
        \addplot table[x=model/snr, y=test/#5, y error=test/#5_std, col sep=comma]{results/perfectsic/#2_#3_rho#6_perfectsic_nd2.csv};
        \label{baselines_#2_#3_rho#6_perfectsic_nd2}
        
        \addplot table[x=model/snr, y=test/#5, y error=test/#5_std, col sep=comma]{results/#2/#3_rho#6_noma_nd16.csv};
        \label{baselines_#2_#3_rho#6_noma_nd16}
        
        \addplot table[x=model/snr, y=test/#5, y error=test/#5_std, col sep=comma]{results/#2/#3_rho#6_noma_nd2.csv};
        \label{baselines_#2_#3_rho#6_noma_nd2}
        
        \addplot table[x=snr, y=cifar10_psnr, col sep=comma]{results_other/nd2_C#6_singlemodel_ft.csv};
        \label{baselines_nd2_C#6_singlemodel_ft}
        
        \addplot table[x=snr, y=cifar10_psnr, col sep=comma]{results_other/nd2_C#6_singlemodel.csv};
        \label{baselines_nd2_C#6_singlemodel}
        
        \addplot table[x=model/snr, y=test/#5, y error=test/#5_std, col sep=comma]{results/#2/#3_rho#6_tdma.csv};
        \label{baselines_#2_#3_rho#6_tdma}
        
        \legend{\scriptsize Perfect SIC (n=16), \scriptsize Perfect SIC (n=2), \scriptsize DeepJSCC-PNOMA (n=16), \scriptsize DeepJSCC-PNOMA (n=2), \scriptsize DeepJSCC-NOMA-CL~\cite{yilmaz2023distributed}, \scriptsize DeepJSCC-NOMA~\cite{yilmaz2023distributed}, \scriptsize DeepJSCC-TDMA}
        \end{axis}
        \end{tikzpicture}
}

\begin{figure*}[t!]
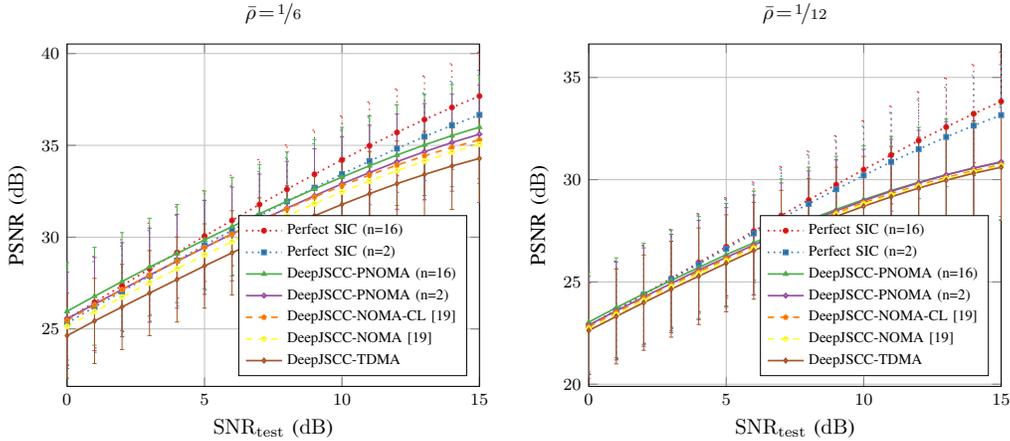

\baselines{$\Bar{\rho}=\nicefrac{1}{6}$}{CIFAR10}{awgn}{PSNR (\si{dB})}{psnr}{6}
\baselines{$\Bar{\rho}=\nicefrac{1}{12}$}{CIFAR10}{awgn}{PSNR (\si{dB})}{psnr}{12}
\caption{Comparison with NOMA-based methods for different bandwidth ratios on CIFAR-10\modified{~\cite{krizhevsky2009learning}}.}
\label{fig:comparison_noma}
\end{figure*}

We employ CIFAR10 and TinyImagenet datasets for comparison, which are described in \cref{sec:datasets}. As benchmark digital coding schemes for comparison, we employ BPG~\cite{Bellard:BPG} and a variety of neural image compression codecs~\cite{cheng2020learned,jiang2023mlicpp,ballé2018variational} in conjunction with 5G LDPC codes for channel encoding. After experimenting with different coding rates and QAM schemes using 5G LDPC codes with a block length of \num{6144} bits, we chose the optimal configuration. A natural extension of DeepJSCC to multi-user case is via time-divison, named DeepJSCC-TDMA, where all the users share the same DeepJSCC encoder and decoder, and each user is only active during its own time slot. We report mean \glspl{PSNR} and standard deviations wherever possible.

\Cref{fig:comparison_point2point} shows the superiority of our method for \(n=16\) compared to the separation-based methods (consisting of BPG or neural codecs combined with LDPC) and to DeepJSCC-TDMA. Our method achieves significantly better reconstruction performance in terms of \(\mathrm{PSNR}\) for all the \(\mathrm{SNR}\) values. Moreover, experiments over a Rayleigh fading channel yield results that are in line with those obtained under the AWGN channel, further demonstrating the robustness of our approach under different channel conditions. This observation also holds true for the \gls{MS-SSIM} and \gls{LPIPS} metrics\modified{, which are defined in the appendix along with SSIM for comprehensive perceptual quality evaluation,} as demonstrated in \appendixOtherMetrics. In \appendixFairness, we further examine the fairness of DeepJSCC-PNOMA among users by evaluating the consistency of their reconstruction quality.

\Cref{fig:comparison_noma} shows the comparison with DeepJSCC-NOMA~\cite{yilmaz2023distributed} and its curriculum learning-based extension DeepJSCC-NOMA-CL~\cite{yilmaz2023distributed}, \modified{and Perfect SIC-based genie-aided model~\cite{yilmaz2023distributed}}. DeepJSCC-NOMA uses NOMA with a shared Siamese encoder and device embeddings for \gls{JSCC}, enabling simultaneous image transmission and reconstruction. Its variant, DeepJSCC-NOMA-CL, first trains on non-interfering signals then fine-tunes on superimposed transmissions to boost robustness and quality. Perfect SIC assumption-based model assumes no interference between users and is decoded only with channel noise and fading. DeepJSCC-PNOMA outperforms DeepJSCC-NOMA and DeepJSCC-NOMA-CL over all \gls{SNR} values. DeepJSCC-PNOMA surprisingly outperforms even the Perfect SIC method when $\mathrm{SNR} < 5 \ $ \si{\decibel} for $\Bar{\rho}=\nicefrac{1}{6}$ and $\mathrm{SNR} < 3 \ $ \si{\decibel} for $\Bar{\rho}=\nicefrac{1}{12}$; implying that orthogonal initialization, training sample subsampling and progressive fine-tuning-based training components are highly effective. We also note that Perfect SIC is an unrealistic assumption since signals naturally interfere with each other in real-world, and particularly in the case of DeepJSCC, it is impossible to decode and cancel interference completely.

\subsection{Correlated Inputs}
\label{sec:comparison_correlated}
\newcommand{\correlatedfigure}[5]{
        \centering
\begin{tikzpicture}[scale=0.8]
        \begin{axis}[
        title={#1},
        xlabel={$\mathrm{SNR}_\mathrm{test}$ (\si{\decibel})},
        error bars/y dir=both,
        error bars/y explicit,
        ylabel={#4},
        cycle list/Set1,
        cycle multiindex* list={
                [2 of]mark list*\nextlist
                Set1\nextlist                {solid,solid,dashed,dashed}\nextlist
        },
        xmin=0,
        xmax=15,
        tick label style={/pgf/number format/fixed}
        ]
        \foreach \bwfactor in {6,12}{
            \addplot table[x=model/snr, y=test/#5, y error=test/#5_std, col sep=comma]{results/#2/#3_rho\bwfactor_noma.csv};
            \addplot table[x=model/snr, y=test/#5, y error=test/#5_std, col sep=comma]{results/#2/#3_rho\bwfactor_tdma.csv};
        }
        \legend{\scriptsize DeepJSCC-PNOMA $(\Bar{\rho}=\nicefrac{1}{6})$,\scriptsize DeepJSCC-TDMA $(\Bar{\rho}=\nicefrac{1}{6})$,\scriptsize DeepJSCC-PNOMA $(\Bar{\rho}=\nicefrac{1}{12})$,\scriptsize DeepJSCC-TDMA $(\Bar{\rho}=\nicefrac{1}{12})$}
        \end{axis}
        \end{tikzpicture}
}

\begin{figure*}[t!]
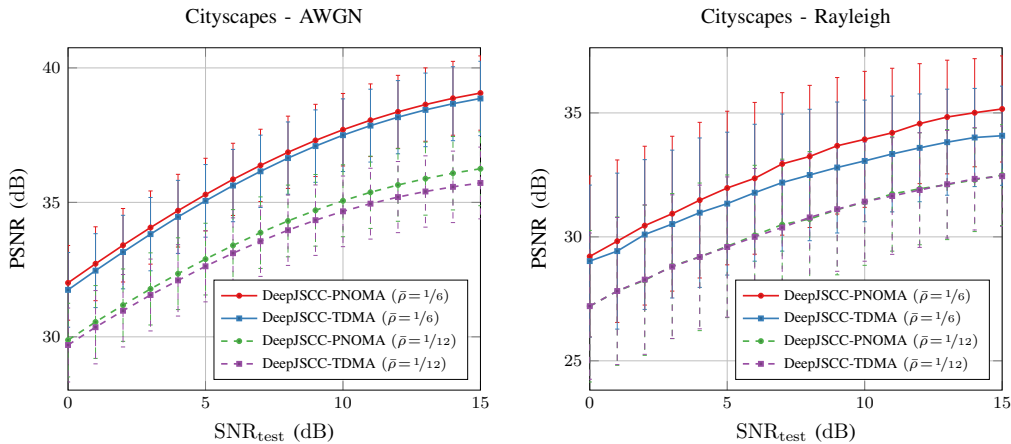

\correlatedfigure{Cityscapes - AWGN}{Cityscape}{awgn}{PSNR (\si{dB})}{psnr}
\correlatedfigure{Cityscapes - Rayleigh}{Cityscape}{rayleigh}{PSNR (\si{dB})}{psnr}
\caption{Performance comparison in terms of \gls{PSNR} on \gls{AWGN} and Rayleigh channels for correlated source images from Cityscapes dataset.}
\label{fig:comparison_correlated}
\end{figure*}

\Cref{fig:comparison_correlated} presents a comparative analysis between DeepJSCC-PNOMA and DeepJSCC-TDMA on the Cityscapes dataset in terms of the average PSNR. The results demonstrate that DeepJSCC-PNOMA consistently outperforms DeepJSCC-TDMA across all evaluated metrics, \glspl{SNR}, and bandwidth ratios. This superior performance indicates the efficacy of our method in leveraging common information between correlated images. By incorporating this correlated information, DeepJSCC-PNOMA is able to more efficiently utilize the available bandwidth, leading to enhanced image reconstruction quality under varying \glspl{SNR}. This observation also holds true for the \gls{MS-SSIM}~\cite{wang2003multiscale} and \gls{LPIPS}~\cite{zhang2018unreasonable} metrics\modified{, detailed in the appendix,} as demonstrated in \appendixCorrelated.

\subsection{Number of Users}
\newcommand{\numusersfigure}[5]{
        \centering
\begin{tikzpicture}[scale=0.5]]
        \begin{axis}[
        title={#1},
        xlabel={$\mathrm{SNR}_\mathrm{test}$ (\si{\decibel})},
        error bars/y dir=both,
        error bars/y explicit,
        ylabel={#4},
        cycle list/Set1-5,
        cycle multiindex* list={
                [5 of]mark list*\nextlist
                Set1-5\nextlist                {solid,solid,solid,solid,solid,dashed,dashed,dashed,dashed,dashed}\nextlist
        },
        xmin=0,
        xmax=15,
        tick label style={/pgf/number format/fixed}
        ]
        \foreach \bwratio in {6,12}{
        \foreach \numusers in {16,8,4,2}{
            \addplot table[x=model/snr, y=test/#5, y error=test/#5_std, col sep=comma]{results/#2/#3_rho\bwratio_noma_nd\numusers.csv};
            \label{numusersfigure_#2_#3_#5_rho\bwratio_noma_nd\numusers}
        }
        \addplot table[x=model/snr, y=test/#5, y error=test/#5_std, col sep=comma]{results/#2/#3_rho\bwratio_tdma.csv};
        \label{numusersfigure_#2_#3_#5_rho\bwratio_noma_nd1}
        }
        \end{axis}
        \end{tikzpicture}
}
\newcommand{\numuserslegend}{
\begin{tabular}{|l|llllll|}
\hline
$\Bar{\rho}=\nicefrac{1}{6}$ &
    \ref*{numusersfigure_CIFAR10_awgn_psnr_rho6_noma_nd16} \multirow{2}{*}{$n=16$} & 
    \ref*{numusersfigure_CIFAR10_awgn_psnr_rho6_noma_nd8} \multirow{2}{*}{$n=8$} &  
    \ref*{numusersfigure_CIFAR10_awgn_psnr_rho6_noma_nd4} \multirow{2}{*}{$n=4$} &  
    \ref*{numusersfigure_CIFAR10_awgn_psnr_rho6_noma_nd2} \multirow{2}{*}{$n=2$} &  
    \ref*{numusersfigure_CIFAR10_awgn_psnr_rho6_noma_nd1} \multirow{2}{*}{$n=1$} & \\
$\Bar{\rho}=\nicefrac{1}{12}$ &
    \ref*{numusersfigure_CIFAR10_awgn_psnr_rho12_noma_nd16} & 
    \ref*{numusersfigure_CIFAR10_awgn_psnr_rho12_noma_nd8} &  
    \ref*{numusersfigure_CIFAR10_awgn_psnr_rho12_noma_nd4} &  
    \ref*{numusersfigure_CIFAR10_awgn_psnr_rho12_noma_nd2} &  
    \ref*{numusersfigure_CIFAR10_awgn_psnr_rho12_noma_nd1} & \\
    \hline
\end{tabular}
}
\begin{figure*}
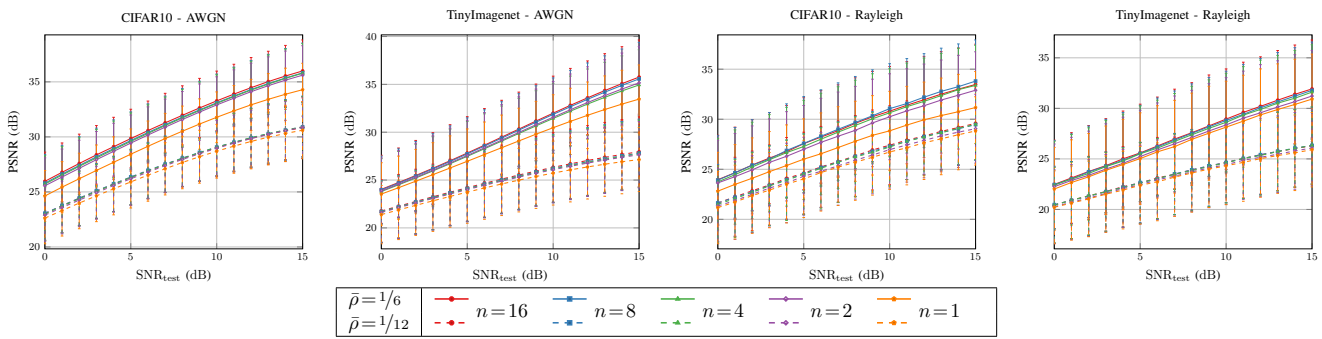

\numusersfigure{CIFAR10 - AWGN}{CIFAR10}{awgn}{PSNR (\si{dB})}{psnr}
\numusersfigure{TinyImagenet - AWGN}{TinyImagenet}{awgn}{PSNR (\si{dB})}{psnr}
\numusersfigure{CIFAR10 - Rayleigh}{CIFAR10}{rayleigh}{PSNR (\si{dB})}{psnr}
\numusersfigure{TinyImagenet - Rayleigh}{TinyImagenet}{rayleigh}{PSNR (\si{dB})}{psnr}
\resizebox{0.5\textwidth}{!}{
\centering
\numuserslegend
}
\caption{Comparison of DeepJSCC-PNOMA with number of users $n \in \LP 1,2,4,8,16 \RP$.}
\label{fig:ablation_numusers}
\end{figure*}
Next, we evaluate our method on CIFAR10 and TinyImagenet datasets for varying numbers of users, $n$. This comparison remains fair by maintaining equal total power and total bandwidth, as described in \cref{remark:fair_comparison}. \Cref{fig:ablation_numusers} illustrates the effect of increasing the number of users on DeepJSCC-PNOMA.
As the number of users grows, the system's performance generally improves due to the potential for more concurrent transmissions. However, this improvement shows diminishing returns, where the incremental gain per additional user decreases. This phenomenon is attributed to increased interference with more users. Consequently, while adding users can initially boost the system performance, the rate of improvement gradually declines as the network nears its operational limits, consistent with the information-theoretic capacity of the \gls{MAC} with respect to $n$.

 \subsection{Model Complexity}
\label{sec:dnn_parameters}
\begin{table}[t!]
\caption{Number of trainable model parameters}
\centering
\begin{tabular}{@{}lrrr@{}}
\toprule
Method      & \# Users   & $\Bar{\rho}=1/6$ & $\Bar{\rho}=1/12$ \\ \midrule
DeepJSCC-TDMA & All &    \num{22200211}  &  \num{22149011}     \\
\midrule
\multirow{5}[-1]{*}{DeepJSCC-PNOMA (Ours)} &  \num{1} & \num{22200211}  &  \num{22149011}     \\
&  \num{2}  &   \num{22202259} & \num{22149523}     \\
 & \num{4} & \num{22208403} & \num{22151059} \\
 & \num{8} & \num{22232979} & \num{22157203} \\
 & \num{16} & \num{22331283} & \num{22181779} \\
 & \modified{\num{32}} & \modified{\num{22724499}} & \modified{\num{22280083}} \\
 & \modified{\num{64}} & \modified{\num{24297363}} & \modified{\num{22673299}} \\
\midrule
DeepJSCC-NOMA (-CL)~\cite{yilmaz2023distributed} & 2 &  \num{22382846}    & \num{22260478}     \\
\midrule
\multirow{5}[-1]{*}{Separate encoders \& decoders} &  \num{1} & \num{22200211}  &  \num{22149011}     \\
 & \num{2} & \num{44402470} & \num{44298534} \\
 & \num{4} & \num{88809036} & \num{88598092} \\
 & \num{8} & \num{177634456} & \num{177200280} \\
 & \num{16} & \num{355334448}  & \num{354416944} \\
 \midrule
 \multirow{2}[-1]{*}{Perfect SIC} & 2 & \num{22322579} & \num{22200211} \\
& 16 & \num{26699667} & \num{23615891} \\
\bottomrule
\end{tabular}
\label{tab:num_parameters}
\end{table}
\Cref{tab:num_parameters} compares the number of trainable parameters for DeepJSCC-TDMA, DeepJSCC-PNOMA, DeepJSCC-NOMA (-CL)~\cite{yilmaz2023distributed}, DeepJSCC-PNOMA without parameter sharing, and the Perfect SIC model over an \gls{AWGN} \gls{MAC}. DeepJSCC-PNOMA with $n=1$ matches DeepJSCC-TDMA in parameters. For $n=2$, DeepJSCC-PNOMA has fewer parameters than DeepJSCC-NOMA and DeepJSCC-NOMA-CL while performing better. Note that DeepJSCC-NOMA (-CL) values are based on CIFAR10, and the parameter counts increase with the resolution of the inputs. \modified{DeepJSCC-PNOMA} scales to $n=16$ users with only $0.6\%$ increase in the number of trainable parameters when $\Bar{\rho}=\nicefrac{1}{6}$ and $0.1\%$ increase when $\Bar{\rho}=\nicefrac{1}{12}$, and further scales to $n=64$ users with $9.4\%$ increase when $\Bar{\rho}=\nicefrac{1}{6}$ and $2.4\%$ increase when $\Bar{\rho}=\nicefrac{1}{12}$. As shown in \cref{remark:scalability}, the number of extra parameters compared to the $n=1$ is $\Oc \LB n^2\Bar{\rho}^2 \RB$.

\modified{Practical implementation challenges such as real-time processing, channel estimation, and mobility are common to all machine learning-based communication systems and are beyond the scope of this work, which focuses on the core algorithmic contributions of DeepJSCC-PNOMA. Nonetheless, our experiments demonstrate \changed{efficient inference with millisecond-level latency}, scalable memory usage, and robust performance across diverse channel conditions. Technical specifications and computational efficiency metrics are detailed in the appendix.}

\changed{Additionally, the network maintains high reconstruction quality while lowering training and inference costs thanks to the progressive expansion and fine-tuning strategy, which reuses previously trained components. The framework is appropriate for deployment in sizable, dynamic communication systems because it demonstrates a well-managed trade-off where incremental complexity permits scalable multi-user performance without sacrificing computational efficiency.}

\subsection{Ablation Study of Introduced Components}
\label{sec:ablation_introduced_components}
\newcommand{\ablationfigure}[6]{
        \centering
\begin{tikzpicture}[scale=0.5]
        \begin{axis}[
        title={#1},
        xlabel={$\mathrm{SNR}_\mathrm{val}$ (\si{\decibel})},
        error bars/y dir=both,
        error bars/y explicit,
        ylabel={#4},
        cycle list/Set1-6,
        cycle multiindex* list={
                [6 of]mark list*\nextlist
                Set1-6\nextlist                {solid,dashed,solid,dashed,dashed,dashed}\nextlist
        },
        xmin=0,
        xmax=15,
        tick label style={/pgf/number format/fixed}
        ]
        \addplot table[x=model/snr, y=test/#5, y error=test/#5_std, col sep=comma]{results/ablation/CIFAR10_awgn_rho#6_ablation_standard_nd16.csv};
        \addplot table[x=model/snr, y=test/#5, y error=test/#5_std, col sep=comma]{results/ablation/CIFAR10_awgn_rho#6_ablation_no_curriculum_learning_nd16.csv};
        \addplot table[x=model/snr, y=test/#5, y error=test/#5_std, col sep=comma]{results/ablation/CIFAR10_awgn_rho#6_ablation_standard.csv};
        \addplot table[x=model/snr, y=test/#5, y error=test/#5_std, col sep=comma]{results/ablation/CIFAR10_awgn_rho#6_ablation_no_curriculum_learning.csv};
        \addplot table[x=model/snr, y=test/#5, y error=test/#5_std, col sep=comma]{results/ablation/CIFAR10_awgn_rho#6_ablation_no_parameter_sharing.csv};
        \addplot table[x=model/snr, y=test/#5, y error=test/#5_std, col sep=comma]{results/ablation/CIFAR10_awgn_rho#6_ablation_no_orthogonal_init.csv};
        \legend{\scriptsize DeepJSCC-PNOMA (n=16),\scriptsize w/o progressive fine-tuning (n=16),\scriptsize DeepJSCC-PNOMA (n=2),\scriptsize w/o progressive fine-tuning (n=2),\scriptsize w/o parameter sharing (n=2),\scriptsize w/o orthogonal initialization (n=2)}
        \end{axis}
        \end{tikzpicture}
}

\newcommand{\ntsablationfigure}[6]{
        \centering
\begin{tikzpicture}[scale=0.5]
        \begin{axis}[
        title={#1},
        xlabel={$\mathrm{SNR}_\mathrm{val}$ (\si{\decibel})},
        error bars/y dir=both,
        error bars/y explicit,
        ylabel={#4},
        cycle list/Set1-5,
        cycle multiindex* list={
                [5 of]mark list*\nextlist
                Set1-5\nextlist                {solid,solid,solid,solid,solid}\nextlist
        },
        xmin=0,
        xmax=15,
        tick label style={/pgf/number format/fixed}
        ]
        \foreach \bwfactor in {#6}{
        \addplot table[x=model/snr, y=test/#5, y error=test/#5_std, col sep=comma]{results/ablation/CIFAR10_awgn_rho\bwfactor_ablation_nts5.csv};
        \addplot table[x=model/snr, y=test/#5, y error=test/#5_std, col sep=comma]{results/ablation/CIFAR10_awgn_rho\bwfactor_ablation_nts4.csv};
        \addplot table[x=model/snr, y=test/#5, y error=test/#5_std, col sep=comma]{results/ablation/CIFAR10_awgn_rho\bwfactor_ablation_standard.csv};
        \addplot table[x=model/snr, y=test/#5, y error=test/#5_std, col sep=comma]{results/ablation/CIFAR10_awgn_rho\bwfactor_ablation_nts2.csv};
        \addplot table[x=model/snr, y=test/#5, y error=test/#5_std, col sep=comma]{results/ablation/CIFAR10_awgn_rho\bwfactor_ablation_nts1.csv};
        }
        \legend{\scriptsize $T = 5N$,\scriptsize $T=4N$,\scriptsize $T=3N$,\scriptsize $T=2N$,\scriptsize $T=N$,}
        \end{axis}
        \end{tikzpicture}
}

\begin{figure*}[t!]
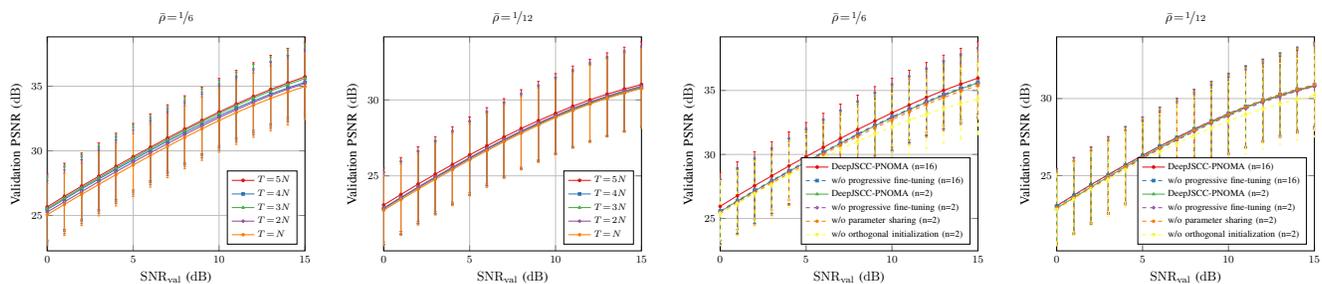

    \ntsablationfigure{$\Bar{\rho}=\nicefrac{1}{6}$}{CIFAR10}{awgn}{Validation PSNR (\si{dB})}{psnr}{6}
\ntsablationfigure{$\Bar{\rho}=\nicefrac{1}{12}$}{CIFAR10}{awgn}{Validation PSNR (\si{dB})}{psnr}{12}
\ablationfigure{$\Bar{\rho}=\nicefrac{1}{6}$}{CIFAR10}{awgn}{Validation PSNR (\si{dB})}{psnr}{6}
\ablationfigure{$\Bar{\rho}=\nicefrac{1}{12}$}{CIFAR10}{awgn}{Validation PSNR (\si{dB})}{psnr}{12}
\caption{Ablation study of the number of the sampled training pairs and the introduced components.}
\label{fig:ablation_components}
\end{figure*}

\Cref{fig:ablation_components} demonstrates the improvements resulting from various components on the validation split of the CIFAR10 dataset to prevent leakage from the test split. The first two plots in \cref{fig:ablation_components} clearly show that increasing the number of training pairs enhances the model's performance when fine-tuning from $n=1$ to $n=2$, albeit with longer training times. The last two plots illustrate the benefits of progressive fine-tuning-based training, orthogonal initialization of user-specific projections, and parameter sharing. Parameter sharing not only boosts performance but also reduces training time. Without progressive fine-tuning, we observe only marginal gains between $n=2$ and $n=16$. However, progressive fine-tuning provides a more stable method for increasing the number of users. Orthogonal initialization is the most effective component of our method, as it ensures that performance does not degrade due to randomness at the start of each fine-tuning step.

\subsection{Orthogonality Analysis}
\label{sec:orthogonality}
\begin{figure*}[t!]
\centering
\includegraphics[width=\textwidth]{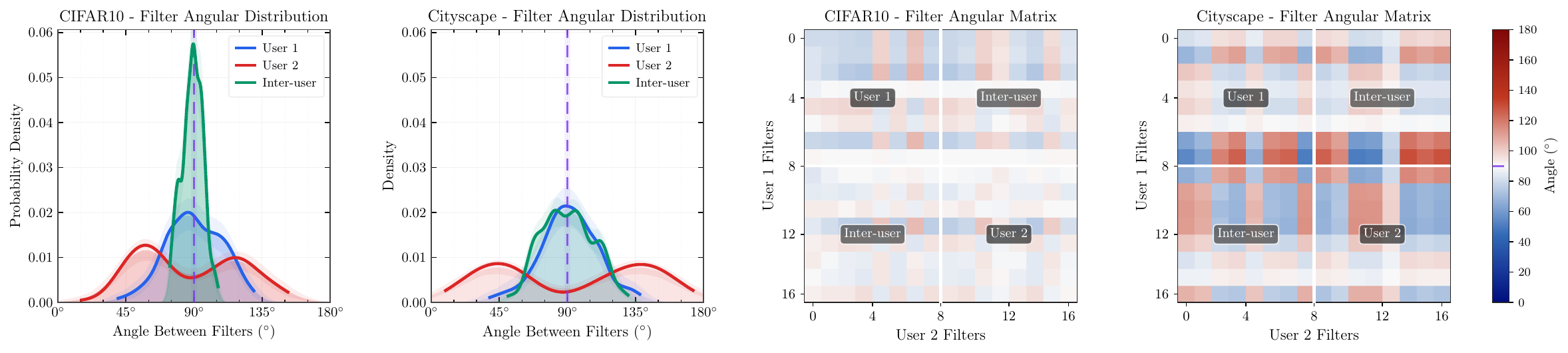}
\caption{Analysis of orthogonality between transmitted filters on Kodak.}
\label{fig:plot_orthogonality_comparison}
\end{figure*}

We examine the orthogonality of encoding filters, which reflect geometric signal separation in DeepJSCC-PNOMA systems. For each user \(u \in \LSB n \RSB\), we define $m$ encoding filters \(\{\mathbf{e}_i^{(u)}\}_{i=1}^{m}\), reshaped into vectors in \(\RR^d\). The angle between any two filters is calculated as:
\begin{equation*}
    \theta_{ij}^{uv} = \arccos\!\left( \frac{\langle \mathbf{e}_i^{(u)}, \mathbf{e}_j^{(v)} \rangle}{\|\mathbf{e}_i^{(u)}\|\|\mathbf{e}_j^{(v)}\|} \right).
    \label{eq:angle}
\end{equation*}
\Cref{fig:plot_orthogonality_comparison} shows filter orthogonality for models trained on independent (CIFAR-10) and correlated (Cityscape) images using two Kodak images transmitted over an \gls{AWGN} channel at \SI{3}{\decibel}. Kernel density estimations (KDEs) indicate angles mostly around $90^{\circ}$, with the model trained on Cityscape dataset displaying greater angular variability, reflecting reduced orthogonality due to shared information between correlated sources. Heatmaps in the third and fourth panels highlight a clear block-diagonal structure distinguishing intra-user from inter-user relationships, resulting from progressive fine-tuning that initially enforces orthogonality. Thus, source correlation critically influences filter orthogonality: independent sources foster near-orthogonality, although perfect orthogonality is not achieved, while correlated sources encourage information sharing by further reducing orthogonal separation.

\subsection{Qualitative Comparison of Reconstructed Outputs and Superposed Signals}
\newcommand{\qualitativerow}[5]{
\vspace{-1em}
\hspace{-.05in}
	\subfigure[$512\times768$] {\includegraphics[width=0.16\textwidth]{figures/comparison_qualitative/#1_original_image.png}}
	\hspace{-.15in}
	\quad
	\subfigure[PSNR (dB)] {\includegraphics[width=0.16\textwidth]{figures/comparison_qualitative/#1_original.png}}
	\hspace{-.20in}
	\quad
	\subfigure[#2]{\includegraphics[width=0.16\textwidth]{figures/comparison_qualitative/#1_bpg_ldpc.png}}
	\hspace{-.20in}
	\quad
	\subfigure[#3] {\includegraphics[width=0.16\textwidth]{figures/comparison_qualitative/#1_cheng2020_ldpc.png}}
	\hspace{-.20in}
	\quad
	\subfigure[#4] {\includegraphics[width=0.16\textwidth]{figures/comparison_qualitative/#1_deepjscc_tdma.png}}
	\hspace{-.20in}
	\quad
	\subfigure[#5]{\includegraphics[width=0.16\textwidth]{figures/comparison_qualitative/#1_deepjscc_pnoma16.png}}
}

\begin{figure*}[!t]
	\begin{subtable}
		\centering
		\scriptsize
		\begin{tabular}{m{0.16\textwidth}m{0.14\textwidth}<{\centering}m{0.14\textwidth}<{\centering}m{0.14\textwidth}<{\centering}m{0.14\textwidth}<{\centering}m{0.14\textwidth}<{\centering}}
			& Original & BPG~\cite{Bellard:BPG} + LDPC & Cheng~\cite{cheng2020learned} + LDPC   &  DeepJSCC-TDMA &  DeepJSCC-PNOMA  \\
		\end{tabular}
	\end{subtable}
	\begin{center}
\qualitativerow{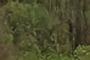}{25.23}{25.37}{25.9}{26.59}
\qualitativerow{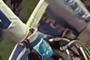}{27.68}{27.78}{28.38}{29.74}
\vspace{2em}
\caption{Qualitative comparison of reconstructed images.}
\label{fig:comparison_qualitative}
\end{center}
\end{figure*}
\begin{figure*}[t!]
    \centering
    \includegraphics[width=\textwidth]{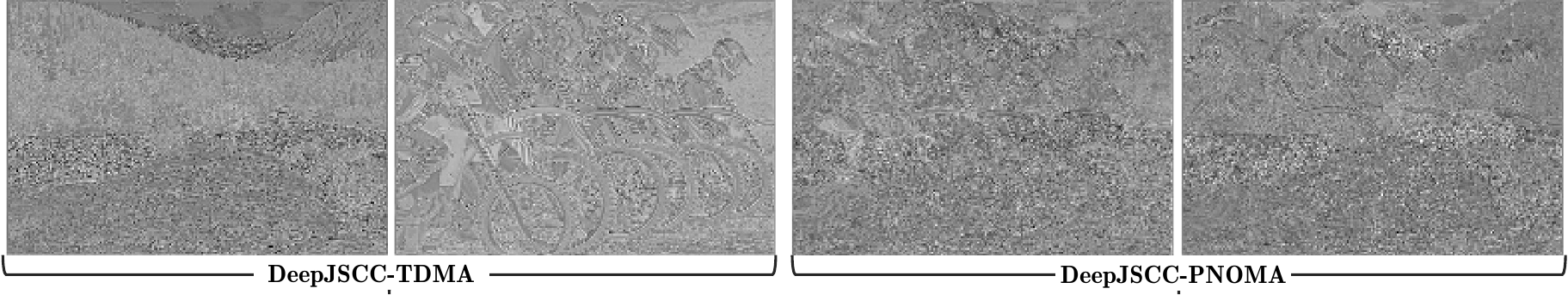}
    \caption{Visualization of transmitted filters for DeepJSCC-TDMA and DeepJSCC-PNOMA.}
    \label{fig:bottlenecks}
\end{figure*}

\Cref{fig:comparison_qualitative} compares the reconstructed images using different coding approaches. DeepJSCC-PNOMA demonstrates significant visual improvement over other methods, validating the quantitative results presented in \cref{sec:comparison_point2point}. \Cref{fig:bottlenecks} visualizes the superposed filters while transmitting the images in \cref{fig:comparison_qualitative} for the two-user case in both DeepJSCC-PNOMA and DeepJSCC-TDMA. In DeepJSCC-TDMA, one user transmits while the other remains silent. In contrast, DeepJSCC-PNOMA allows both users to transmit simultaneously, making both of their images discernible from the superposed filters. 

\subsection{Analysis of Losses}
\label{sec:validation_losses}
\newcommand{\lossesfigure}[6]{
        \centering
\begin{tikzpicture}[scale=0.8]
        \begin{axis}[
        cycle list/Set1-6,
        cycle multiindex* list={
                [6 of]mark list*\nextlist
                Set1-6\nextlist                {solid,dashed,solid,dashed,dashed,dashed}\nextlist
        },
        x tick scale label style={yshift=4pt},
            title={#1},
            ymode=log,
            xlabel={Training Step},
            ylabel={#4},
            legend style={
                at={(0.98,0.98)}, 
                anchor=north east,
                font=\scriptsize,
                legend columns=1
            },
            grid=both,
            major grid style={line width=.1pt,draw=gray!30},
            title style={font=\small},
            label style={font=\small},
            tick label style={font=\scriptsize},
            every axis plot/.append style={line width=0.7pt},
        ]
        \addplot table[x=step, y=#5, col sep=comma]{results/losses_ablation/CIFAR10_awgn_rho#6_ablation_standard_nd16.csv};
        \addplot table[x=step, y=#5, col sep=comma]{results/losses_ablation/CIFAR10_awgn_rho#6_ablation_no_curriculum_learning_nd16.csv};
        \addplot table[x=step, y=#5, col sep=comma]{results/losses_ablation/CIFAR10_awgn_rho#6_ablation_standard.csv};
        \addplot table[x=step, y=#5, col sep=comma]{results/losses_ablation/CIFAR10_awgn_rho#6_ablation_no_curriculum_learning.csv};
        \addplot table[x=step, y=#5, col sep=comma]{results/losses_ablation/CIFAR10_awgn_rho#6_ablation_no_orthogonal_init.csv};
        \addplot table[x=step, y=#5, col sep=comma]{results/losses_ablation/CIFAR10_awgn_rho#6_ablation_tdma.csv};
        \legend{\scriptsize DeepJSCC-PNOMA (n=16),\scriptsize w/o progressive fine-tuning (n=16),\scriptsize DeepJSCC-PNOMA (n=2),\scriptsize w/o progressive fine-tuning (n=2),\scriptsize w/o orthogonal initialization (n=2),\scriptsize DeepJSCC-TDMA}
        \end{axis}
        \end{tikzpicture}
}
\begin{figure*}[t!]
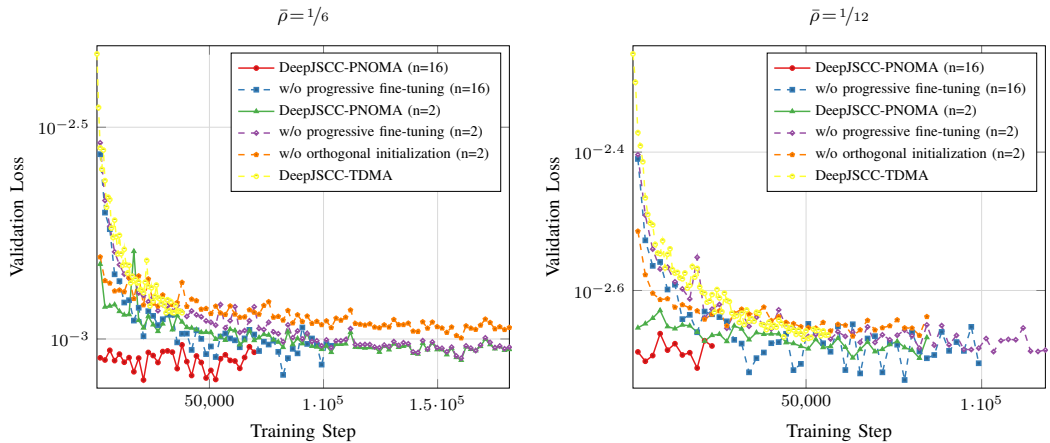

\lossesfigure{$\Bar{\rho}=\nicefrac{1}{6}$}{CIFAR10}{awgn}{Validation Loss}{val/loss}{6}
\lossesfigure{$\Bar{\rho}=\nicefrac{1}{12}$}{CIFAR10}{awgn}{Validation Loss}{val/loss}{12}
\caption{Visualization of losses throughout the training or fine-tuning steps.}
\label{fig:losses_ablation}
\end{figure*}

\Cref{fig:losses_ablation} illustrates the validation losses throughout the training process for bandwidth ratios $\Bar{\rho}=\nicefrac{1}{6}$ and $\Bar{\rho}=\nicefrac{1}{12}$. In both plots, the DeepJSCC-PNOMA model employing progressive fine-tuning exhibits a notably higher initial loss and converges to a higher final loss for both \( n = 2 \) and \( n = 16 \). This discrepancy is particularly pronounced at \( \Bar{\rho} = \nicefrac{1}{6} \). Similarly, the absence of orthogonal initialization results in elevated initial and final losses, potentially due to the model being trapped in a local optimum, likely caused by the interference from other users. These differences can probably be attributed to the random weights leading to increased interference from other users during the fine-tuning process. As observed with DeepJSCC-TDMA (equivalent to DeepJSCC-PNOMA with $n=1$), DeepJSCC-PNOMA with $n=2$, and DeepJSCC-PNOMA with $n=16$, the final loss value decreases as the number of users $n$ increases, aligning with the expected capacity gains of multiple access channels with more users.

\subsection{Scalability to Higher User Counts}
\label{sec:scalability_higher_users}
\newcommand{\comparisonTableCaption}{\modified{Scalability analysis: Performance comparison on CIFAR-10 under AWGN channel for different bandwidth ratios. Values shown as PSNR (dB). \textbf{Bold values} indicate best performance within each bandwidth ratio group.}}

\newcommand{\comparisonTableContent}{%
  \small
  \begin{tabular}{@{}lcc*{4}{c}@{}}
    \toprule
    & & \multicolumn{4}{c}{\textbf{SNR (dB)}} \\
    \cmidrule(lr){3-6}
    \textbf{Bandwidth} & \textbf{Users} & \textbf{0} & \textbf{3} & \textbf{7} & \textbf{10} \\
    \textbf{Ratio} & & & & & \\
    \midrule
    \multirow{3}{*}{$\Bar{\rho} = 1/6$} 
    & $n=16$ & 25.96 & 28.37 & 31.28 & 33.27 \\
    & $n=32$ & 26.03 & 28.42 & 31.33 & \textbf{33.32} \\
    & $n=64$ & \textbf{26.05} & \textbf{28.46} & \textbf{31.34} & 33.29 \\
    \addlinespace[0.3em]
    \multirow{3}{*}{$\Bar{\rho} = 1/12$} 
    & $n=16$ & 23.04 & 25.07 & 27.49 & 29.03 \\
    & $n=32$ & 23.18 & 25.18 & \textbf{27.57} & \textbf{29.08} \\
    & $n=64$ & \textbf{23.20} & \textbf{25.19} & 27.53 & 28.98 \\
    \bottomrule
  \end{tabular}%
}

\ifdefined\tworeviewresponse
  \begin{center}
  \captionof{table}{\comparisonTableCaption}
  \label{tab:comparison_16_32_64}
  \comparisonTableContent
  \end{center}
\else
  \begin{table}[tbp]
    \centering
    \caption{\comparisonTableCaption}
    \label{tab:comparison_16_32_64}
    \comparisonTableContent
  \end{table}
\fi

\modified{To further validate the scalability and generalizability of our approach, we extend our experiments to systems with 32 and 64 users. \Cref{tab:comparison_16_32_64} presents a comprehensive performance comparison across different user counts (16, 32, and 64) under AWGN channel conditions for both $\Bar{\rho} = 1/6$ and $\Bar{\rho} = 1/12$ power constraints.}

\modified{The results demonstrate that DeepJSCC-PNOMA maintains consistent performance as the number of users scales from 16 to 64, with marginal improvements in PSNR and LPIPS metrics. Notably, the system shows stable performance across all SNR levels tested (0, 3, 7, and 10 dB), indicating robust scalability. For $\Bar{\rho} = 1/6$, PSNR improvements range from 0.07-0.09 dB when scaling from 16 to 64 users, while for $\Bar{\rho} = 1/12$, improvements range from 0.16-0.19 dB. The LPIPS metric shows similar stability with minimal variations across different user counts.}

\modified{These results confirm that our progressive fine-tuning approach and orthogonal initialization strategy effectively handle the increased interference and complexity associated with larger user populations, making the proposed method viable for large-scale deployments.}

\changed{Beyond image transmission, a variety of network architectures and dynamic scenarios can benefit from the core concepts of our work, including mobile and vehicular communications in 5G/6G systems, IoT sensor networks with time-varying device populations, and drone swarm coordination with rapidly changing topologies. The progressive user expansion and orthogonal projection design allow adaptive scaling as users join or leave the network, while the fine-tuning mechanisms could enable lightweight online updates under time-varying channels or user mobility. Although exploring such real-time adaptations lies beyond the current scope, the proposed framework provides a strong foundation for scalable and low-latency deep joint source–channel coding in next-generation communication networks.}

\section{Conclusion}
\label{sec:conclusion}
We have introduced a novel joint image compression and transmission scheme for multi-user uplink scenarios, leveraging \gls{NOMA} with identical \gls{DNN}-based encoders and decoders for all users. Utilizing a user-specific projection trick, inspired by the CDMA scheme, the receiver can recover images from multiple users despite the analog transmission inherent in DeepJSCC, correctly attributing each image to its respective user. Our DeepJSCC-PNOMA scheme outperforms digital and DeepJSCC-based point-to-point alternatives. Furthermore, it scales up to $64$ users with only an extra $9.4\%$ of trainable parameters at $\Bar{\rho}=\nicefrac{1}{6}$ and $2.4\%$ at $\Bar{\rho}=\nicefrac{1}{12}$, demonstrating consistent performance gains.

\bibliographystyle{IEEEtran}
\bibliography{output}

\cleardoublepage
\appendix

\newcommand{\secCurriculumLearning}{Section~IV-D}
\newcommand{\secSubsampling}{Section~IV-C}
\newcommand{\secSystemModel}{Section~III}
\newcommand{\secMethodology}{Section~IV}
\newcommand{\secComparisonPointToPoint}{Section~V-C}
\newcommand{\secOrthogonality}{Section~V-H}
\newcommand{\secValidationLosses}{Section~V-J}
\newcommand{\secComparisonCorrelated}{Section~V-D}
\newcommand{\secDatasets}{Section~V-A}

\subsection{Additional Details of DeepJSCC-PNOMA}
\label{sec:appendix_additional_methodology}

\subsubsection{Construction of Training and Evaluation samples}
\label{sec:appendix_construction}
\modified{\Cref{alg:construction_train,alg:construction_eval} outlines the method used to construct the training and evaluation samples, respectively, which are detailed in \secSubsampling.}

\tikzexternaldisable
\begin{algorithm}
\caption{\strut Construction of training samples}
\label{alg:construction_train}
\begin{algorithmic}
    \LComment{Inputs: number of tuples T, size of training data $N=\card{\Dctrain}$, number of users $n$}
    \LComment{Generate a vector of random permutation of $nT$ indices}
    \State $\tv = \mathrm{RandomPermutation} \LB \LSB 1,2,\ldots,nT \RSB^\mathrm{T} \RB$
    \LComment{Match them to the training data size}
    \State $\tv = \tv \,\, \mathrm{mod} \,\, n$
    \LComment{Assign indices to users by reshaping}
    \State $\Tm = \mathrm{Reshape}\LB \tv, \LB T, n \RB \RB$
    \LComment{$\Tm$ can now be used as indices of the samples for training, where first dimension is shuffled at every epoch}
    \end{algorithmic}
\end{algorithm}
\tikzexternalenable

\tikzexternaldisable
\begin{algorithm}
\caption{\strut Construction of evaluation (validation and test) samples}
\label{alg:construction_eval}
\begin{algorithmic}
    \LComment{Inputs: number of users $n$, size of evaluation data $M$ (either the cardinality of validation data $\card{\Dcval}$ or the cardinality of test data $\card{\Dctest}$)}
    \LComment{Generate a vector of random permutation of $M$ indices}
    \State $\tv = \mathrm{RandomPermutation} \LB \LSB 1,2,\ldots,M \RSB^\mathrm{T} \RB$
    \LComment{Assign indices to users by reshaping}
    \State $\Tm = \mathrm{Reshape} \LB \tv, \LB \frac{M}{n}, n \RB \RB$
    \LComment{$\Tm$ can now be used as indices of the samples for evaluation}
    \end{algorithmic}
\end{algorithm}
\tikzexternalenable

\subsection{Technical Specifications and Computational Efficiency}
\label{sec:appendix_technical_specifications}


\modified{During evaluation, the model processes inference efficiently with sub-second latency for reconstruction across all user counts tested. The progressive fine-tuning strategy contributes to training stability while maintaining reasonable computational overhead compared to training from scratch for each user configuration.}

\modified{Memory usage remains manageable even for larger user counts due to the shared encoder-decoder architecture and the parameter sharing mechanism described in \secMethodology. The system's ability to scale from 1 to 64 users while maintaining consistent performance demonstrates the practical viability of the proposed approach for real-world multi-user communication scenarios.}

\begin{table*}[t!]
\caption{Training durations and number of epochs passed before early stopping. Duration format is hours:minutes:seconds.}
\centering
\begin{tabular}{@{}lrrrr@{}}
\toprule
                                                   & \multicolumn{2}{c}{Training Duration} & \multicolumn{2}{c}{Number of Epochs} \\ \cmidrule(l){2-5} 
Method &
  \multicolumn{1}{c}{$\Bar{\rho}=\nicefrac{1}{6}$} &
  \multicolumn{1}{c}{$\Bar{\rho}=\nicefrac{1}{12}$} &
  \multicolumn{1}{c}{$\Bar{\rho}=\nicefrac{1}{6}$} &
  \multicolumn{1}{c}{$\Bar{\rho}=\nicefrac{1}{12}$} \\ \midrule
DeepJSCC-PNOMA ($n=1$)                               & 1:04:38            & 1:50:16          & 54                & 80               \\
DeepJSCC-PNOMA ($n=2$)                               & 4:42:27            & 2:09:58          & 86                & 40               \\
DeepJSCC-PNOMA ($n=4$)                               & 3:49:57            & 2:34:43          & 42                & 28               \\
DeepJSCC-PNOMA ($n=8$)                               & 10:41:58           & 7:14:31          & 31                & 21               \\
DeepJSCC-PNOMA ($n=16$)                              & 20:24:06           & 3:40:45          & 33                & 11               \\ \midrule
DeepJSCC-PNOMA ($n=2$)                               & 4:42:27            & 2:09:58          & 86                & 40               \\
DeepJSCC-PNOMA w/o progressive fine-tuning ($n=2$)   & 8:12:00            & 7:22:08          & 86                & 56               \\
DeepJSCC-PNOMA w/o orthogonal initialization ($n=2$) & 8:09:30            & 5:55:50          & 86                & 40               \\
DeepJSCC-PNOMA w/o parameter sharing ($n=2$)         & 5:43:02            & 7:58:39          & 86                & 59               \\ \bottomrule
\end{tabular}
\label{tab:durations_train}
\end{table*}

\subsubsection{Further Dataset Details}
\label{sec:appendix_datasets}

\modified{\textbf{Dataset Overview:} Our experimental evaluation employs four diverse datasets to comprehensively assess the performance and generalizability of DeepJSCC-PNOMA across different image characteristics, resolutions, and application domains. This multi-dataset approach ensures robust validation of our method's effectiveness across varied visual content types and complexity levels.}

\textbf{Dataset Licenses:} All these datasets are publicly accessible under permissible licenses. 
We use CIFAR10 dataset by following the MIT License; TinyImagenet by following the MIT license since it is a subset of Imagenet; Cityscapes by following specific license agreement on its website (\url{https://www.cityscapes-dataset.com/license/}) that is permissive and publicly available for academic purposes; and Kodak by following GNU GPLv3 license.

\modified{\textbf{Dataset Characteristics:}
\begin{itemize}
    \item \textbf{CIFAR-10}\modified{~\cite{krizhevsky2009learning}}: Contains 60,000 32×32 color images across 10 classes (50,000 training, 10,000 test). Selected for its standardized low-resolution format, enabling efficient experimentation and comparison with existing literature.
    \item \textbf{TinyImageNet}: Comprises 200 classes with 64×64 color images (100,000 training, 10,000 validation, 10,000 test). Provides intermediate resolution complexity between CIFAR-10 and full ImageNet, testing scalability across image sizes.
    \item \textbf{Cityscapes}\modified{~\cite{cordts2016cityscapes}}: Urban scene dataset with 1024×2048 high-resolution images, focusing on semantic urban understanding. Used specifically for correlated input experiments due to spatial and semantic relationships between urban scenes.
    \item \textbf{Kodak}: Professional photography dataset with 24 high-quality 512×768 images, commonly used in image compression research. Provides evaluation on natural, high-fidelity photographic content with diverse textures and colors.
\end{itemize}}

\textbf{Dataset Sources:} We use CIFAR10 downloaded by Torchvision; TinyImagenet dataset downloaded from \url{https://github.com/rmccorm4/Tiny-Imagenet-200}; Cityscapes dataset downloaded from \url{https://www.cityscapes-dataset.com/}; Kodak dataset downloaded from \url{https://huggingface.co/datasets/Freed-Wu/kodak}.

\modified{\textbf{Preprocessing and Augmentation:} All datasets undergo consistent preprocessing: normalization to [0,1] range, random horizontal flips during training (except Kodak), and appropriate resizing when necessary. No additional augmentation techniques are applied to maintain fair comparison with baseline methods and preserve the original image characteristics essential for transmission quality evaluation.}

\modified{\textbf{Train/Validation/Test Splits:} For CIFAR-10 and TinyImageNet, we use standard splits provided by the datasets. For Cityscapes, we utilize the official train/validation split for correlated experiments. For Kodak, given its small size, we use all 24 images for evaluation purposes in fairness analysis, consistent with established image compression evaluation protocols.}

\subsubsection{Hardware Requirements}
We conduct all our deep learning experiments by training models with the distributed data parallel method on an internal cluster setup, featuring 2 x NVIDIA RTX A6000 GPUs, each with 48GB of GPU memory, and an Intel(R) Core(TM) i9-10980XE CPU. The same Intel i9-10980XE CPU is also utilized for data compression with the BPG method.

\subsubsection{Runtime and Memory Discussion}

We present the training durations, conducted in a shared, non-optimal environment with multiple processes and an uneven workload.

\textbf{Training Durations:}
\Cref{tab:durations_train} presents the training durations and the number of epochs completed before early stopping for a subset of the trained models. CIFAR10 training durations range from $2$ to $20$ hours. TinyImageNet training durations range from $5$ to $80$ hours. Cityscapes training durations range from $45$ to $160$ minutes. Training durations mainly depend on the number of tuples $T$, per-user bandwidth \modified{$\Bar{\rho}$}, training dataset size $\Dctrain$, and the number of users $n$. The table highlights the effectiveness of techniques such as progressive fine-tuning, orthogonal initialization, and parameter sharing in accelerating the training process. As the number of users increases, these techniques become even more critical, facilitating faster model convergence in terms of epochs and reducing the computational burden associated with training. This not only enhances the performance of the DeepJSCC-PNOMA model but also ensures its efficiency and scalability across diverse multi-user communication environments. However, the increase in training duration with a higher number of users underscores the need for further computational optimizations, such as quantization and model pruning. Additionally, it is important to note that some training durations vary significantly due to the uneven workload on the servers during the training process.

\textbf{Evaluation Durations:} All evaluations at a given SNR are completed in under one minute, with minimal variation, primarily influenced by the size of the validation and test datasets. 

\textbf{Memory:} For evaluations, all experiments require less than $3$ GB of GPU memory. For CIFAR10 training, approximately $1.5n$ GB of GPU memory is utilized per GPU. For TinyImageNet training, around $2.4$ GB of GPU memory is used per GPU. For Cityscapes training, $12$ GB of GPU memory is used for $n=1$, and $32$ GB of GPU memory is used for $n=2$.

\subsubsection{Software Requirements}
The code to reproduce experiments requires the following software dependencies: Python $3.9.0$ or higher, PyTorch (torch) version $2.1.0$ or higher, Torchvision version $0.16.0$ or higher, Lightning version $2.0.6$, and Torchmetrics version $1.3.1$. PyTorch provides a platform for building and training neural networks, Torchvision offers datasets and model architectures for computer vision, Lightning standardizes high-performance deep learning research, and Torchmetrics supplies performance metrics for model evaluation. Once Python and PyTorch are installed manually, all the necessary dependencies can be installed by running \lstinline[language=bash]|pip install -r requirements.txt| in the main directory.

\subsection{Additional Experiments}
\label{sec:appendix_additional_experiments}

\subsubsection{Analysis of Fairness}
\label{sec:appendix_fairness}
\newcommand{\fairnessfigure}[2]{
\begin{tikzpicture}[scale=0.5]
        \begin{axis}[
        title={#1},
        xlabel={$\mathrm{SNR}_\mathrm{test}$ (\si{\decibel})},
        error bars/y dir=both,
        error bars/y explicit,
        ylabel={PSNR (\si{dB})},
        cycle list/Set1-5,
        cycle multiindex* list={
                [5 of]mark list*\nextlist
                Set1-5\nextlist
                {dashed,dashed,dashed,dashed,solid}\nextlist
        },
        xmin=0,
        xmax=15
        ]
            \addplot table[x=snr, y=psnr_user1, col sep=comma]{results/fairness/Kodak_#2_rho6_nd4.csv};
            \addplot table[x=snr, y=psnr_user2, col sep=comma]{results/fairness/Kodak_#2_rho6_nd4.csv};
            \addplot table[x=snr, y=psnr_user3, col sep=comma]{results/fairness/Kodak_#2_rho6_nd4.csv};
            \addplot table[x=snr, y=psnr_user4, col sep=comma]{results/fairness/Kodak_#2_rho6_nd4.csv};
            \addplot table[x=snr, y=tdma, col sep=comma]{results/fairness/Kodak_#2_rho6_nd4.csv};
        \legend{$1^\mathrm{st}$ user,  $2^\mathrm{nd}$ user,  $3^\mathrm{rd}$ user,  $4^\mathrm{th}$ user,DeepJSCC-TDMA}
        \end{axis}
        \end{tikzpicture}
}
\newcommand{\fairnessscatter}[1]{
\begin{tikzpicture}[scale=0.5]
\begin{axis}[
  xlabel = $1^{\mathrm{st}}$ user's average PSNR (\si{dB}),
  ylabel = $2^{\mathrm{nd}}$ user's average PSNR (\si{dB}),
    cycle list/Set1-3,
    cycle multiindex* list={
            [3 of]mark list*\nextlist
            Set1-3\nextlist
            {solid,solid,solid}\nextlist
    },
  enlarge y limits=0,
  mark size=3.5pt,
  title = {#1}]
  \addplot table[only marks, scatter,x=psnr_user1,y=psnr_user2,col sep=comma] {results/fairness/scatter_Kodak_awgn_rho6_snr0.csv};
  \addplot table[only marks, scatter,x=psnr_user1,y=psnr_user2,col sep=comma] {results/fairness/scatter_Kodak_awgn_rho12_snr0.csv};
  \addplot [no marks,domain=22:33.5] {x};
  \legend{$\Bar{\rho}=\nicefrac{1}{6}$, $\Bar{\rho}=\nicefrac{1}{12}$, $y=x$}
\end{axis}
\end{tikzpicture}
}
\begin{figure}[t!]
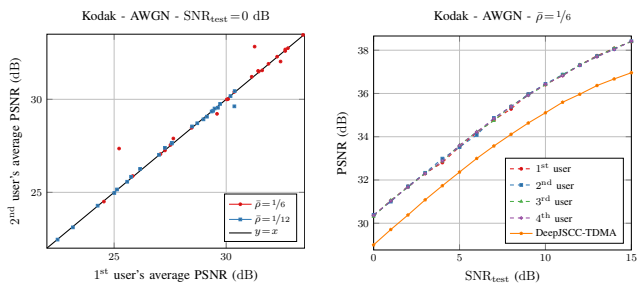

\centering
\fairnessscatter{Kodak - AWGN - $\mathrm{SNR}_\mathrm{test}=0$ \si{dB}}
\fairnessfigure{Kodak - AWGN - $\Bar{\rho}=\nicefrac{1}{6}$}{awgn}   
\caption{Analysis of fairness between users on Kodak dataset.}
\label{fig:plot_fairness_comparison}
\end{figure}

In this subsection, we analyze the fairness of DeepJSCC-PNOMA, ensuring that all users obtain comparable image reconstruction quality even under varying channel conditions—a critical attribute in multi-user communication systems. We formally define the fairness objective as follows:

\begin{definition}[Fairness Objective]
\label{def:fairness}
Let \(\text{PSNR}_i\) denote the average Peak Signal-to-Noise Ratio (PSNR) for user \(i\). The system is considered fair if the differences \(|\text{PSNR}_i - \text{PSNR}_j|\) are negligible for all \(i,j \in \LSB n \RSB\).
\end{definition}

To evaluate this metric, we conducted experiments on the Kodak dataset using a DeepJSCC-PNOMA model with \(n=4\) users for the first plot and \(n=2\) users for the second plot, both trained on CIFAR10 dataset on \gls{AWGN} channel.  All plots in \cref{fig:plot_fairness_comparison} are derived from experiments on the Kodak dataset. \Cref{fig:plot_fairness_comparison} shows the fairness analysis for the DeepJSCC-PNOMA method. The first plot presents the PSNR values for two users transmitting the same image under a fixed channel condition (\(\mathrm{SNR}=0 \ \si{\decibel}\)); the nearly identical PSNR values confirm that the system maintains fairness even under challenging noise conditions. The second plot, which displays PSNR performance across a range of SNR values, further demonstrates that the system consistently delivers uniform image reconstruction quality regardless of channel variations. Together, these results robustly validate our fairness criterion and underscore the capability of DeepJSCC-PNOMA to provide an equitable quality of service across different users.

\subsubsection{Additional Evaluation Metrics}
\begin{figure*}
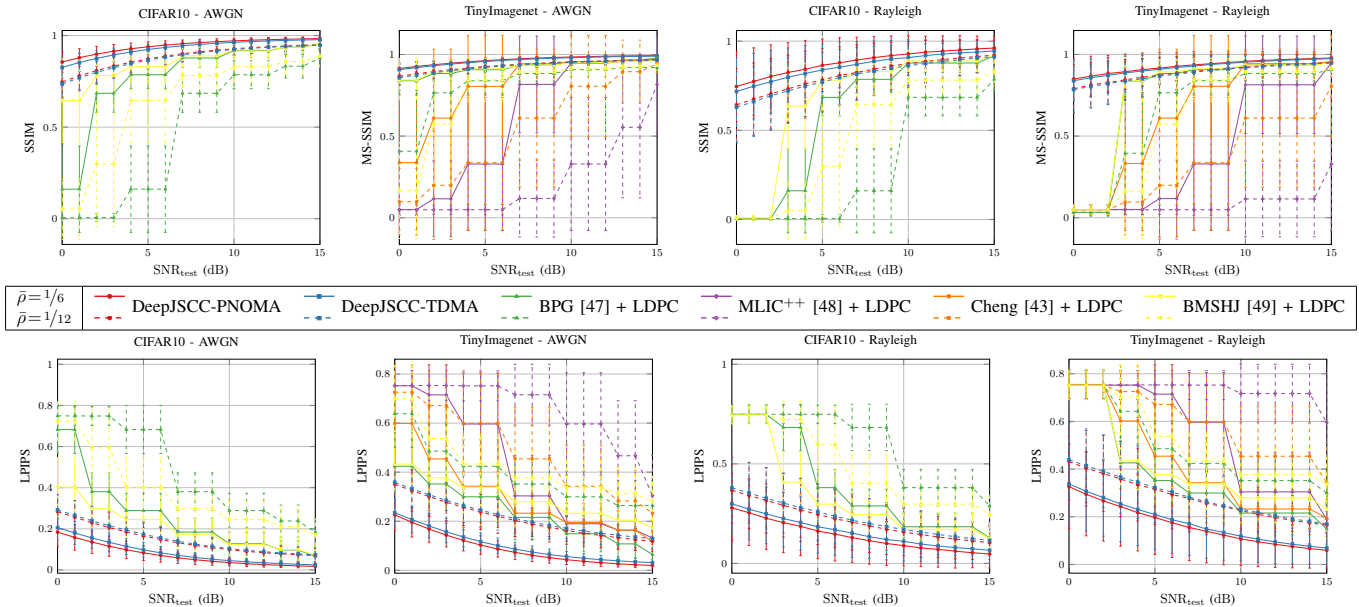

\separationawgnfigureCIFAR{CIFAR10 - AWGN}{CIFAR10}{awgn}{SSIM}{ssim}{ldpc}
\separationawgnfigure{TinyImagenet - AWGN}{TinyImagenet}{awgn}{MS-SSIM}{msssim}{ldpc}
\separationawgnfigureCIFAR{CIFAR10 - Rayleigh}{CIFAR10}{rayleigh}{SSIM}{ssim}{ldpc}
\separationawgnfigure{TinyImagenet - Rayleigh}{TinyImagenet}{rayleigh}{MS-SSIM}{msssim}{ldpc}
\resizebox{\textwidth}{!}{
\begin{tabular}{|l|llllll|}
\hline
$\Bar{\rho}=\nicefrac{1}{6}$ &
  \ref*{separationawgnfigure_TinyImagenet_awgn_msssim_rho6_noma_nd16_ldpc}
  \multirow{2}{*}{DeepJSCC-PNOMA} &
  \ref*{separationawgnfigure_TinyImagenet_awgn_msssim_rho6_tdma_ldpc}
  \multirow{2}{*}{DeepJSCC-TDMA} &
  \ref*{separationawgnfigure_TinyImagenet_awgn_msssim_rho6_bpg_ldpc}
  \multirow{2}{*}{BPG~\cite{Bellard:BPG} + LDPC} &
  \ref*{separationawgnfigure_TinyImagenet_awgn_msssim_rho6_mlicpp_ldpc}
  \multirow{2}{*}{MLIC$^{++}$~\cite{jiang2023mlicpp} + LDPC} &
  \ref*{separationawgnfigure_TinyImagenet_awgn_msssim_rho6_cheng2020_anchor_ldpc}
  \multirow{2}{*}{Cheng~\cite{cheng2020learned} + LDPC} &
  \ref*{separationawgnfigure_TinyImagenet_awgn_msssim_rho6_bmshj2018_ldpc}
  \multirow{2}{*}{BMSHJ~\cite{ballé2018variational} + LDPC}  \\
$\Bar{\rho}=\nicefrac{1}{12}$ &
  \ref*{separationawgnfigure_TinyImagenet_awgn_msssim_rho12_noma_nd16_ldpc} &
  \ref*{separationawgnfigure_TinyImagenet_awgn_msssim_rho12_tdma_ldpc} &
  \ref*{separationawgnfigure_TinyImagenet_awgn_msssim_rho12_bpg_ldpc} &
  \ref*{separationawgnfigure_TinyImagenet_awgn_msssim_rho12_mlicpp_ldpc} &
  \ref*{separationawgnfigure_TinyImagenet_awgn_msssim_rho12_cheng2020_anchor_ldpc} &
  \ref*{separationawgnfigure_TinyImagenet_awgn_msssim_rho12_bmshj2018_ldpc}
   \\ \hline
\end{tabular}
}
\separationawgnfigureCIFAR{CIFAR10 - AWGN}{CIFAR10}{awgn}{LPIPS}{lpips}{ldpc}
\separationawgnfigure{TinyImagenet - AWGN}{TinyImagenet}{awgn}{LPIPS}{lpips}{ldpc}
\separationawgnfigureCIFAR{CIFAR10 - Rayleigh}{CIFAR10}{rayleigh}{LPIPS}{lpips}{ldpc}
\separationawgnfigure{TinyImagenet - Rayleigh}{TinyImagenet}{rayleigh}{LPIPS}{lpips}{ldpc}
\caption{Comparison with point-to-point DeepJSCC and separation-based methods over SSIM, MS-SSIM and LPIPS metrics.}
\label{fig:comparison_point2point_othermetrics}
\end{figure*}
In addition to \gls{PSNR} metric defined in \modified{\secSystemModel}, we also perform comparisons for \gls{SSIM} , \gls{MS-SSIM} and \gls{LPIPS}. For our evaluations, we use SSIM for CIFAR10 due to its resolution, MS-SSIM for TinyImagenet, and LPIPS for both datasets.

The Structural Similarity Index (SSIM) between two images \(\vec{x}\) and \(\hat{\vec{x}}\) is defined as:
\[
\text{SSIM}(\vec{x}, \hat{\vec{x}}) = \frac{(2\mu_{\vec{x}}\mu_{\hat{\vec{x}}} + c_1)(2\sigma_{\vec{x}\hat{\vec{x}}} + c_2)}{(\mu_{\vec{x}}^2 + \mu_{\hat{\vec{x}}}^2 + c_1)(\sigma_{\vec{x}}^2 + \sigma_{\hat{\vec{x}}}^2 + c_2)},
\]
where \(\mu_{\vec{x}}\) is the mean of image \(\vec{x}\), \(\mu_{\hat{\vec{x}}}\) is the mean of image \(\hat{\vec{x}}\), \(\sigma_{\vec{x}}^2\) is the variance of image \(\vec{x}\), \(\sigma_{\hat{\vec{x}}}^2\) is the variance of image \(\hat{\vec{x}}\), \(\sigma_{\vec{x}\hat{\vec{x}}}\) is the covariance between images \(\vec{x}\) and \(\hat{\vec{x}}\), \(c_1\) and \(c_2\) are constants to stabilize the division with weak denominators.

The \gls{MS-SSIM} metric extends the \gls{SSIM} by evaluating image quality at multiple scales. \gls{MS-SSIM} involves computing \gls{SSIM} at different scales (usually created by iteratively downsampling the images) and combining these measurements into a single score. The images are iteratively downsampled to create a series of images at different scales. At each scale $j$, the SSIM index is computed for the downsampled images. These SSIM scores are then combined using a set of weights 
$w_{j}$ for each scale.
\[
\text{MS-SSIM}(\vec{x}, \hat{\vec{x}}) = \left[ \prod_{j=1}^M \text{SSIM}(\vec{x}^j, \hat{\vec{x}}^j)^{w_j} \right]^{\frac{1}{\sum_{j=1}^M w_j}},
\]
where \(M\) is the number of scales, \(\vec{x}^j\) and \(\hat{\vec{x}}^j\) are the images \(\vec{x}\) and \(\hat{\vec{x}}\) at scale \(j\), \(\text{SSIM}(\vec{x}^j, \hat{\vec{x}}^j)\) is the SSIM index at scale \(j\), \(w_j\) are the weights for each scale \(j\), for which we use default values by the original paper except the filter size that is chosen as the maximum possible value according to the image resolution that is lower than or equal to the default value $11$~\cite{wang2003multiscale}. After calculating this metric for every image in the dataset, we take average over the images in the dataset. \gls{MS-SSIM} has been shown to perform better at representing human perception compared to \gls{PSNR}.
\Gls{LPIPS} is a perception metric~\cite{zhang2018unreasonable}, which computes the similarity between the activations of two image patches for a pretrained neural network, such as VGG or AlexNet. Lower \gls{LPIPS} scores indicate greater perceptual similarity between the patches. The \gls{LPIPS} metric has become popular in image processing tasks like image super-resolution, GAN evaluation, and other applications where perceptual quality is crucial. It is valued for its ability to better reflect human visual perception compared to traditional pixel-based metrics. We employ pretrained VGG network to evaluate \gls{LPIPS}.

\subsubsection{Comparison on SSIM, MS-SSIM and LPIPS Metrics}
\label{sec:appendix_comparison_othermetrics}

\cref{fig:comparison_point2point_othermetrics} shows the comparison of our method with separation-based alternatives combined with LDPC codes. These results align with our discussion in \secComparisonPointToPoint, despite being evaluated using different metrics. This consistency supports the generalization of our method to achieve high perceptual quality.

\begin{figure*}
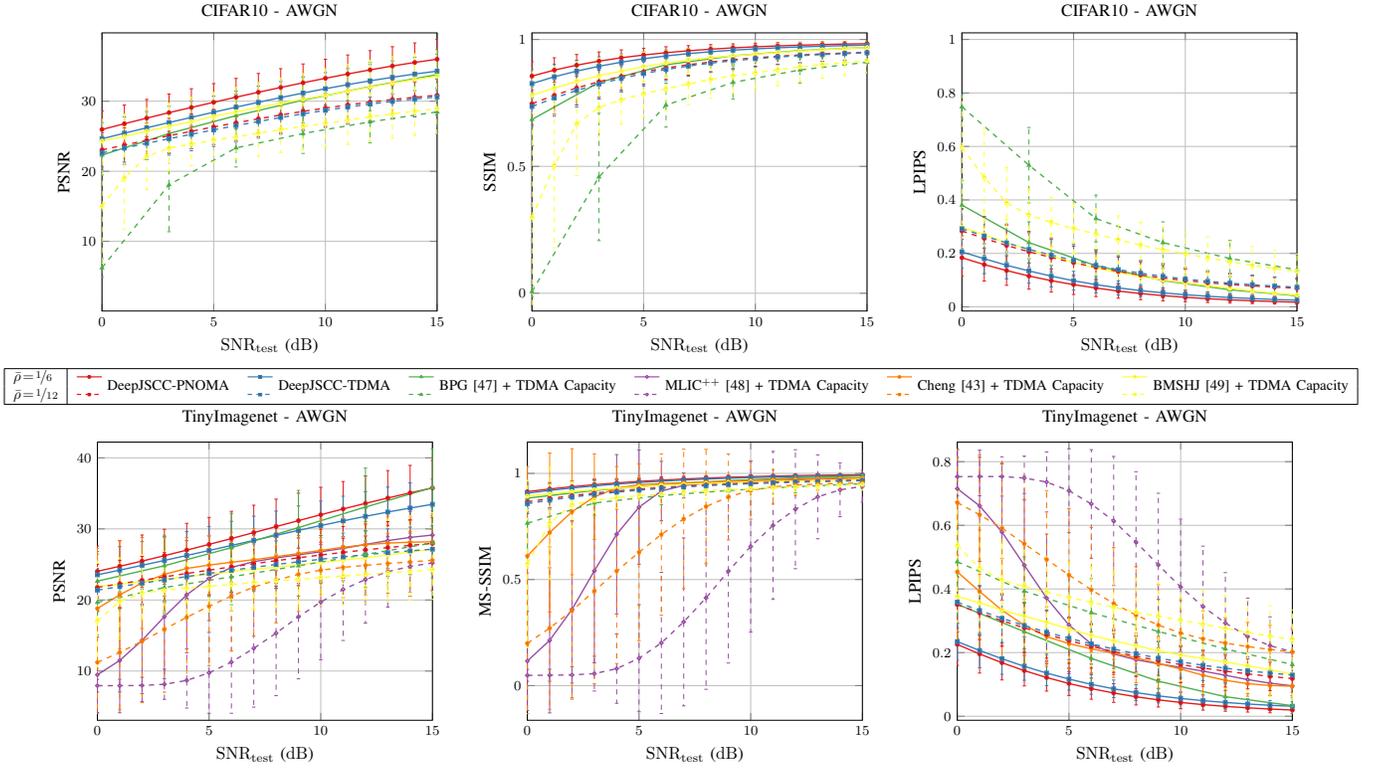

\separationawgnfigureCIFARbig{CIFAR10 - AWGN}{CIFAR10}{awgn}{PSNR}{psnr}{capacity1}
\separationawgnfigureCIFARbig{CIFAR10 - AWGN}{CIFAR10}{awgn}{SSIM}{ssim}{capacity1}
\separationawgnfigureCIFARbig{CIFAR10 - AWGN}{CIFAR10}{awgn}{LPIPS}{lpips}{capacity1}
\resizebox{\textwidth}{!}{
\begin{tabular}{|l|llllll|}
\hline
$\Bar{\rho}=\nicefrac{1}{6}$ &
  \ref*{separationawgnfigure_TinyImagenet_awgn_msssim_rho6_noma_nd16_ldpc}
  \multirow{2}{*}{DeepJSCC-PNOMA} &
  \ref*{separationawgnfigure_TinyImagenet_awgn_msssim_rho6_tdma_ldpc}
  \multirow{2}{*}{DeepJSCC-TDMA} &
  \ref*{separationawgnfigure_TinyImagenet_awgn_msssim_rho6_bpg_ldpc}
  \multirow{2}{*}{BPG~\cite{Bellard:BPG} + TDMA Capacity} &
  \ref*{separationawgnfigure_TinyImagenet_awgn_msssim_rho6_mlicpp_ldpc}
  \multirow{2}{*}{MLIC$^{++}$~\cite{jiang2023mlicpp} + TDMA Capacity} &
  \ref*{separationawgnfigure_TinyImagenet_awgn_msssim_rho6_cheng2020_anchor_ldpc}
  \multirow{2}{*}{Cheng~\cite{cheng2020learned} + TDMA Capacity} &
  \ref*{separationawgnfigure_TinyImagenet_awgn_msssim_rho6_bmshj2018_ldpc}
  \multirow{2}{*}{BMSHJ~\cite{ballé2018variational} + TDMA Capacity}  \\
$\Bar{\rho}=\nicefrac{1}{12}$ &
  \ref*{separationawgnfigure_TinyImagenet_awgn_msssim_rho12_noma_nd16_ldpc} &
  \ref*{separationawgnfigure_TinyImagenet_awgn_msssim_rho12_tdma_ldpc} &
  \ref*{separationawgnfigure_TinyImagenet_awgn_msssim_rho12_bpg_ldpc} &
  \ref*{separationawgnfigure_TinyImagenet_awgn_msssim_rho12_mlicpp_ldpc} &
  \ref*{separationawgnfigure_TinyImagenet_awgn_msssim_rho12_cheng2020_anchor_ldpc} &
  \ref*{separationawgnfigure_TinyImagenet_awgn_msssim_rho12_bmshj2018_ldpc}
   \\ \hline
\end{tabular}
}
\separationawgnfigurebig{TinyImagenet - AWGN}{TinyImagenet}{awgn}{PSNR}{psnr}{capacity1}
\separationawgnfigurebig{TinyImagenet - AWGN}{TinyImagenet}{awgn}{MS-SSIM}{msssim}{capacity1}
\separationawgnfigurebig{TinyImagenet - AWGN}{TinyImagenet}{awgn}{LPIPS}{lpips}{capacity1}
\caption{Comparison with point-to-point DeepJSCC and separation-based methods with capacity over PSNR, SSIM, MS-SSIM and LPIPS metrics.}
\label{fig:comparison_point2point_othermetrics_capacity1}
\end{figure*}

\Cref{fig:comparison_point2point_othermetrics_capacity1} compares our method on the AWGN MAC with separation-based alternatives and capacity on CIFAR10 and TinyImageNet datasets. Achieving capacity is highly optimistic and generally not feasible in the real world. We do not use any specific practical channel coding or modulation schemes to determine this bound. Compressing the source at the maximum possible rate and assuming error-free transmission requires a capacity-achieving combination of channel coding and modulation for reliable transmission. Therefore, the performance of any separation-based transmission scheme using an actual channel coding scheme and modulation with BPG compression will likely fall short of this upper bound. Despite this, our method performs significantly better in all evaluated SNRs. We also note that in low-SNR regime separation-based alternatives often fail to transmit as seen in the figure.

\subsubsection{Additional Comparison on Correlated Inputs}
\label{sec:appendix_comparison_correlated}

\begin{figure*}
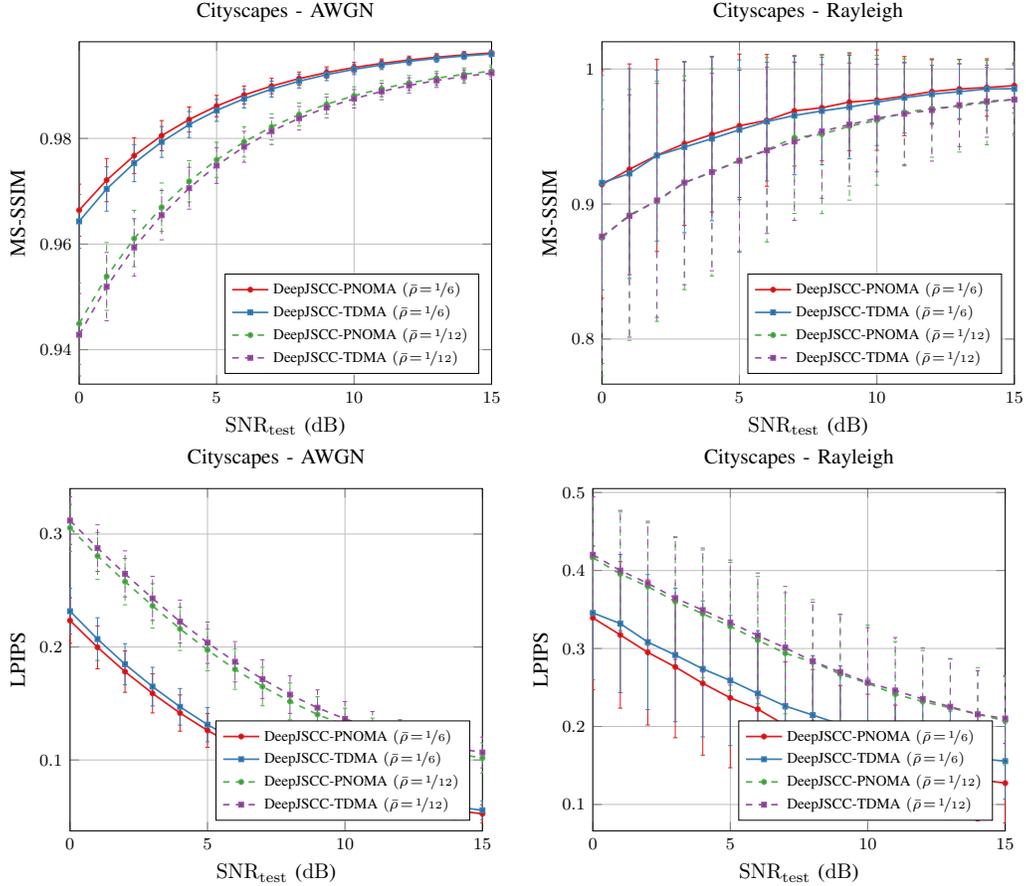

\correlatedfigure{Cityscapes - AWGN}{Cityscape}{awgn}{MS-SSIM}{msssim}
\correlatedfigure{Cityscapes - Rayleigh}{Cityscape}{rayleigh}{MS-SSIM}{msssim}
\correlatedfigure{Cityscapes - AWGN}{Cityscape}{awgn}{LPIPS}{lpips}
\correlatedfigure{Cityscapes - Rayleigh}{Cityscape}{rayleigh}{LPIPS}{lpips}
\caption{Performance comparison in terms of \gls{MS-SSIM} and \gls{LPIPS} on \gls{AWGN} and Rayleigh channels for correlated source images from Cityscapes dataset.}
\label{fig:comparison_correlated_appendix}
\end{figure*}

\Cref{fig:comparison_correlated_appendix} presents a comparative analysis between DeepJSCC-PNOMA and DeepJSCC-TDMA on the Cityscapes dataset for MS-SSIM and LPIPS metrics. Similar to PSNR, for MS-SSIM and LPIPS, DeepJSCC-PNOMA obtains the best performance for all evaluated bandwidth ratios and \glspl{SNR}.

\subsubsection{Comparison for Varying Number of Actively Participating Users}
\label{sec:appendix_comprehensiveness}

\newcommand{\comprehensivenessfigure}[5]{
        \centering
\begin{tikzpicture}[scale=0.5]]
        \begin{axis}[
        title={#1},
        xlabel={$\mathrm{SNR}_\mathrm{test}$ (\si{\decibel})},
        ylabel={#4},
        cycle list/Set1-5,
        cycle multiindex* list={
                [5 of]mark list*\nextlist
                Set1-5\nextlist                {solid,solid,solid,solid,solid,dashed,dashed,dashed,dashed,dashed}\nextlist
        },
        xmin=0,
        xmax=15,
        tick label style={/pgf/number format/fixed}
        ]
        \foreach \bwratio in {6,12}{
        \foreach \numusers in {16,8,4,2,1}{
            \addplot table[x=model/snr, y=test/#5, y error=test/#5_std, col sep=comma]{results/comprehensiveness/comprehensiveness_#2_#3_rho\bwratio_noma_np\numusers.csv};
            \label{comprehensivenessfigure_#2_#3_#5_rho\bwratio_noma_nd\numusers}
        }
        }
        \end{axis}
        \end{tikzpicture}
}
\newcommand{\numactiveuserslegend}{
\begin{tabular}{|l|llllll|}
\hline
$\Bar{\rho}=\nicefrac{1}{6}$ &
    \ref*{comprehensivenessfigure_CIFAR10_awgn_psnr_rho6_noma_nd16} \multirow{2}{*}{$\card{\Pc}=16$} & 
    \ref*{comprehensivenessfigure_CIFAR10_awgn_psnr_rho6_noma_nd8} \multirow{2}{*}{$\card{\Pc}=8$} &  
    \ref*{comprehensivenessfigure_CIFAR10_awgn_psnr_rho6_noma_nd4} \multirow{2}{*}{$\card{\Pc}=4$} &  
    \ref*{comprehensivenessfigure_CIFAR10_awgn_psnr_rho6_noma_nd2} \multirow{2}{*}{$\card{\Pc}=2$} &  
    \ref*{comprehensivenessfigure_CIFAR10_awgn_psnr_rho6_noma_nd1} \multirow{2}{*}{$\card{\Pc}=1$} & \\
$\Bar{\rho}=\nicefrac{1}{12}$ &
    \ref*{comprehensivenessfigure_CIFAR10_awgn_psnr_rho12_noma_nd16} & 
    \ref*{comprehensivenessfigure_CIFAR10_awgn_psnr_rho12_noma_nd8} &  
    \ref*{comprehensivenessfigure_CIFAR10_awgn_psnr_rho12_noma_nd4} &  
    \ref*{comprehensivenessfigure_CIFAR10_awgn_psnr_rho12_noma_nd2} &  
    \ref*{comprehensivenessfigure_CIFAR10_awgn_psnr_rho12_noma_nd1} & \\
    \hline
\end{tabular}
}
\begin{figure*}
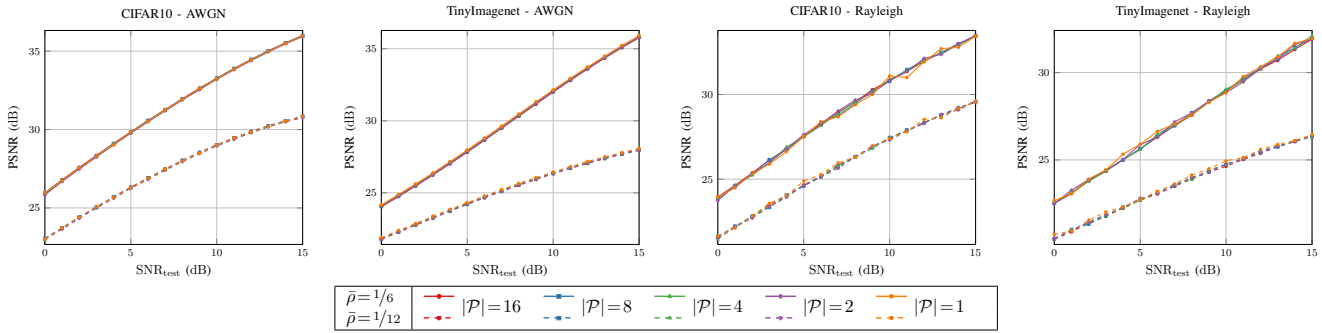

\comprehensivenessfigure{CIFAR10 - AWGN}{CIFAR10}{awgn}{PSNR (\si{dB})}{psnr}
\comprehensivenessfigure{TinyImagenet - AWGN}{TinyImagenet}{awgn}{PSNR (\si{dB})}{psnr}
\comprehensivenessfigure{CIFAR10 - Rayleigh}{CIFAR10}{rayleigh}{PSNR (\si{dB})}{psnr}
\comprehensivenessfigure{TinyImagenet - Rayleigh}{TinyImagenet}{rayleigh}{PSNR (\si{dB})}{psnr}
\resizebox{0.5\textwidth}{!}{
\centering
\numactiveuserslegend
}
\caption{Comparison of DeepJSCC-PNOMA for number of active users $\card{\Pc} \in \LP 1,2,4,8,16 \RP$ to demonstrate the comprehensiveness of our method according to the \cref{def:comprehensiveness}.}
\label{fig:ablation_comprehensiveness}
\end{figure*}

\begin{definition}[Comprehensiveness Objective]
\label{def:comprehensiveness}
    Let \(\text{PSNR}_\Pc\) be the average \gls{PSNR} over the whole test set when only a subset of users $\Pc \subset \LSB n \RSB$ transmit, i.e., $\zv_i=\mathbf{0}, \forall i \not\in \Pc$. The system is said to be comprehensive if $\forall \Pc \subset \LSB n \RSB$, $\text{PSNR}_\Pc \geq \text{PSNR}_{\LSB n \RSB}$.
\end{definition}

\Cref{fig:ablation_comprehensiveness} compares the performance of DeepJSCC-PNOMA with \( n = 16 \) users against scenarios with fewer actively transmitting users, specifically for \( \Pc \in \{1, 2, 4, 8, 16\} \). This comparison enables us to evaluate the comprehensiveness objective outlined in \cref{def:comprehensiveness}. 

Notably, even though the DeepJSCC network for \( n = 16 \) is not explicitly trained for scenarios with fewer actively transmitting users, the performance remains remarkably consistent. This consistency is likely attributable to the near-orthogonality of the user signals, as discussed in \modified{\secOrthogonality}.

These results clearly demonstrate that the method can be effectively used with a variable number of actively transmitting users, aligning with the comprehensiveness objective across most channel conditions. We also posit that there is a potential performance gain resulting from the reduced interference when fewer users are transmitting. This phenomenon can be explained by the increase in information-theoretic capacity under such conditions. In our experiments, we did not observe this effect because non-transmitting users were not included during the training phase. We leave the inclusion of non-transmitting users for future work.

\end{document}